\documentclass[%
 aip,
 jcp,
 amsmath,amssymb,
 reprint,%
]{revtex4-1}

\usepackage{graphicx}%
\usepackage{dcolumn}%
\usepackage{bm}%

\usepackage[utf8]{inputenc}
\usepackage[T1]{fontenc}
\usepackage{mathptmx}
\usepackage{etoolbox}
\usepackage[breaklinks=true,colorlinks,citecolor=blue,linkcolor=blue,urlcolor=blue]{hyperref}
\usepackage{cleveref}
\usepackage{booktabs} 
\usepackage{soul}
\usepackage{bbm}
\usepackage{amsmath,amssymb}
\usepackage{comment}
\usepackage{tikz}
\usetikzlibrary{positioning,arrows.meta,calc}

\definecolor{stepcol}{RGB}{40,40,40}
\definecolor{sdftcol}{RGB}{53, 202, 191}
\definecolor{scdftcol}{RGB}{168, 44, 53}
\definecolor{mggacol}{RGB}{131, 66, 164}

\tikzset{
  box/.style={draw=stepcol, rounded corners=3pt, thick, align=center,
              font=\Large, inner sep=5pt, fill=white},
  diffbox/.style={box, draw=sdftcol, very thick, fill=sdftcol!8},
  flow/.style={-{Latex[length=2.6mm]}, thick, stepcol},
  panelname/.style={font=\bfseries\huge, anchor=west},
  note/.style={font=\Large, align=center},
}

\newcommand{\rv}{\mathbf{r}}
\newcommand{\sg}{\sigma\sigma'}

\newcommand{\MnIr}{Mn$_3$Ir}
\newcommand{\NiS}{NiS$_2$}
\newcommand{\MnGe}{Mn$_3$Ge}
\newcommand{\YMO}{YMnO$_3$}

\newcommand{\diff}{\mathrm{d}}
\newcommand{\muB}{\mu_\mathrm{B}}
\newcommand{\imag}{\text{i}}

\usepackage[normalem]{ulem}
\usepackage{xcolor}

\makeatletter
\def\@email#1#2{%
 \endgroup
 \patchcmd{\titleblock@produce}
  {\frontmatter@RRAPformat}
  {\frontmatter@RRAPformat{\produce@RRAP{*#1\href{mailto:#2}{#2}}}\frontmatter@RRAPformat}
  {}{}
}
\makeatother
\begin{document}

\preprint{AIP/123-QED}

\title{A test drive for exchange-correlation functionals on noncollinear magnets: Mn$_3$Ir, Mn$_3$Ge, NiS$_2$, and YMnO$_3$}
\author{Marie-Therese Huebsch}
\affiliation{VASP Software GmbH, Berggasse 21/14, A-1090 Vienna, Austria}
\email{marie-therese.huebsch@vasp.at}
\author{Martijn Marsman}
\affiliation{VASP Software GmbH, Berggasse 21/14, A-1090 Vienna, Austria}
\author{Jacques K. Desmarais}
\affiliation{Dipartimento di Chimica, Università di Torino, via Giuria 5, 10125 Torino, Italy}
\author{Stefano Pittalis}
\affiliation{Istituto Nanoscienze – CNR, S3, Via Campi 213A, I-41125 Modena, Italy}
\author{Fabien Tran}
\affiliation{VASP Software GmbH, Berggasse 21/14, A-1090 Vienna, Austria}

\date{\today}

\begin{abstract}
A first-principles description of noncollinear magnets is challenging, and the practical behavior of genuinely noncollinear exchange-correlation approximations is largely untested. Spin-current density-functional theory
(SCDFT) admits noncollinear approximations constrained by local U(1)$\times$SU(2) gauge invariance, an exact condition unavailable in the more restrictive framework of spin-DFT (SDFT).
Using the Vienna ab initio simulation package (VASP), we compare two recently developed noncollinear SCDFT functionals, NCMSCAN and LFNCBR89-NCCS, against locally collinear SDFT extensions of four common functionals (PZ, PW92, PBE, and SCAN) for several noncollinear $d$-electron systems.
The key finding is that among all tested functionals, only the SCDFT functionals recover the spin texture of the experimental ground state of NiS$_2$, and they do so at a cost comparable to standard semi-local approximations. Yet the two SCDFT functionals differ in equilibrium volumes, band gaps, and on-site magnetic moments.
Overall, this survey illustrates the usefulness of SCDFT in capturing noncollinear physics with the tested approximations.
\end{abstract}

\maketitle

\section{Introduction}
\label{sec:Introduction}

Every practical prediction of density-functional theory (DFT)\cite{Hohenberg1964,Kohn1965,marzari2021electronic} stands or falls with the exchange-correlation (XC) approximation.\cite{Becke2014:JCP:18,Burke2012:JCP:150901,Capelle2006:BJP:1318,tran2026semi} Developing such approximations for noncollinear magnets is especially challenging\cite{sandratskii1998noncollinear,goings2018current} because the delicate balance between exchange and correlation controls not only structural and electronic properties or the magnitudes of magnetic moments,\cite{Fu2018applicability,Ekholm2018assessing,Jana2018:JCP:044120,MejiaRodriguez2019:PRB:041113,Tran2020magnetism} but also the relative stability of competing magnetic orders.\cite{blugel1988ferromagnetism,horton2019high,huebsch2021benchmark,huebsch2022magnetic,nomoto2024high,nomoto2026systematic} Unsurprisingly, the modern noncollinear XC approximations we will employ draw significant influence from earlier works of Becke on collinear functionals: the expansion of the exchange hole \cite{becke1983hartree} and its generalization to the inclusion of currents,\cite{Becke1996:CJC:995,Becke2002:JCP:6935} the electron localization function,\cite{becke1990simple} and the Becke--Roussel exchange-hole model.\cite{becke1989exchange}

Most noncollinear magnetic first-principles calculations\cite{sandratskii1998noncollinear,marzari2021electronic} use spin-DFT (SDFT),\cite{von1972local} which usually involves a scheme of projecting noncollinear densities onto a local spin-quantization axis.\cite{kubler1988density,kubler1988local,oda1998fully,Hobbs2000:PRB:11556,van2002spin,kurz2004ab,corso2005spin,torrent2008implementation,tancogne2017self} 
The projection is done in order to employ XC functionals that take spin-up and spin-down electron densities as basic variables. 
This prescription, originally developed for local functionals, was heuristically extended to (meta-)generalized gradient approximations, (M)GGAs, making them \emph{locally collinear}.\cite{Hobbs2000:PRB:11556,peralta2007noncollinear} Other schemes of embedding collinear semi-local functionals into noncollinear SDFT exist,\cite{van2002spin,scalmani2012new,pu2023noncollinear,Franzke2025:JCP:084104} which are not considered in this work.

Spin-current DFT (SCDFT)\cite{Vignale1987:PRL:2360,vignale1988current,bencheikh2003spin}
employs noncollinear semi-local XC functionals\cite{Abedinpour2010:PRB:125123,Huebsch2026:PRB:105101,Desmarais2025:PRL:106402} with the added benefit that they obey exact constraints that are unavailable within the SDFT approach.\cite{Abedinpour2010:PRB:125123,Pittalis2017:PRB:035141} Namely, SCDFT functionals take $2\times2$ densities including
the paramagnetic charge-current and spin-current densities as basic variables and thus the XC energy is invariant under U(1)$\times$SU(2) gauge transformations.
Two of these functionals, NCMSCAN\cite{Desmarais2025:PRL:106402} and LFNCBR89-NCCS,\cite{Huebsch2026:PRB:105101} are available in the Vienna ab initio simulation package (VASP),\cite{kresse1993ab,kresse1994ab} but practical experience with both remains limited to proof-of-principle applications.
Theoretically, SCDFT provides a consistent treatment of vector-potential couplings, spin currents, and spin-orbit coupling (SOC).\cite{Vignale1987:PRL:2360,vignale1988current,bencheikh2003spin}
The price of admitting charge and spin-currents is paid both in implementation complexity and in functional design.\cite{Abedinpour2010:PRB:125123,Pittalis2017:PRB:035141,DesmaraisPRL2024,Huebsch2026:PRB:105101,Desmarais2025:PRL:106402,desmarais2020adiabatic,bodo2022spin,comaskey2022role}

NCMSCAN, the noncollinear version of the modified strongly constrained and appropriately normed (MSCAN) functional,\cite{Desmarais2025:PRL:106402} was constructed by enforcing the U(1)$\times$SU(2) invariance, yet through a minimal modification of the SCAN functional.\cite{sun2015strongly} Coincidentally, it also tames the well-documented\cite{Fu2018applicability,Ekholm2018assessing,Jana2018:JCP:044120,MejiaRodriguez2019:PRB:041113,Tran2020magnetism} overmagnetization of SCAN. The functional was first implemented in the development version of the Gaussian-basis ab initio code package CRYSTAL.\cite{erba2022crystal23}

LFNCBR89-NCCS, the functional formulated by Huebsch, Tran, and Marsman,\cite{Huebsch2026:PRB:105101} combines a Laplacian-free (LF) noncollinear variant (NC) of the Becke--Roussel exchange functional\cite{becke1989exchange,Pittalis2017:PRB:035141,tancogne2023constructing} (BR89) with a noncollinear form of the Colle--Salvetti correlation expression\cite{colle1975approximate,Pittalis2007:JCP:124103,tancogne2023constructing,tancogne2023constructing_erratum} (CS).
BR89, whose construction starts from a model for the exchange hole, has subsequently been used in different contexts, most notably,
for calculating the dipole moment in a dispersion-interaction model\cite{Becke2005:JCP:154101} or approximating the exact exchange potential,\cite{Becke2006:JCP:221101} noting that the latter work triggered follow-up studies.\cite{Tran2009:PRL:226401,Raesaenen2010:JCP:44112}
CS is best known as the parent expression of the Lee--Yang--Parr (LYP) functional for correlation.\cite{Lee1988lyp,Miehlich1989:CPL:200} LYP has been combined mostly with the B88 exchange functional of Becke\cite{Becke1988:PRA:3098} to form either a GGA (BLYP) or a hybrid (B3LYP\cite{Becke1993:JCP:5648,Stephens1994:JPC:11623}), which have been very popular in the chemistry community.\cite{Sousa2007:JPCA:10439,casida2026axeldieterbeckejust}

The aforementioned noncollinear SCDFT functionals inherit much of their parameter content from collinear ancestors.\cite{tran2026semi} It is not obvious that these functionals will simultaneously improve on predictions for the properties one usually cares about, e.g., equilibrium volumes, electronic band gaps, and on-site magnetic moments, and capture the qualitatively new physics they were designed for. The declared target for these functionals is noncollinear magnetism including the magnetic order, the description of spin-orbit torques, magnetic anisotropy and, by extension, transition temperatures, and magnetic transport effects. 
Only the application to real noncollinear magnetic materials and a comparison against experimental references can decide whether these functionals have a competitive advantage over established locally collinear functionals.

Here, we compare NCMSCAN and LFNCBR89-NCCS to locally collinear SDFT extensions of four collinear parent functionals: the Perdew--Zunger (PZ)\cite{perdew1981self} and Perdew--Wang (PW92)\cite{Perdew1992:PRB:13244} LSDAs, the Perdew--Burke--Ernzerhof (PBE) GGA,\cite{perdew1996generalized} and the SCAN\cite{sun2015strongly} MGGA.
In our survey we are focusing on a small set of $d$-electron antiferromagnetic (AFM) noncollinear magnets, namely the metals \MnIr\ (cubic) and \MnGe\ (hexagonal), the narrow-gap Mott insulator \NiS\ (cubic), and the multiferroic insulator \YMO\ (hexagonal). 

The manuscript is organized as follows: In Sec.~\ref{sec:Methods}, we discuss the locally collinear SDFT and noncollinear SCDFT self-consistent field (SCF) loops (Sec.~\ref{subsec:ele min}). We introduce the tested XC approximations (Sec.~\ref{subsec:XC functionals}) and provide details on the workflows. Finally, we present a comparison of the functional-resolved computational costs (Sec.~\ref{subsec:Computational cost}). In Sec.~\ref{sec:Results}, we present volume relaxations for each material and XC functional to compare the equilibrium lattice constants and bulk moduli (Sec.~\ref{subsec:Structural properties}), as well as fundamental band gaps and on-site magnetic moments (Sec.~\ref{subsec:moments and gaps}) to low-temperature experimental references (Table~\ref{tab:exp lattice}) where they exist. We then map out the magnetic energy landscape over symmetry-classified magnetic configurations for \NiS\ and \MnIr\ (Sec.~\ref{subsec:Magnetic energy landscape}). For \NiS, we include an on-site Coulomb interaction $U$ and analyze the functional dependence of the band dispersion (Sec.~\ref{subsec:band-structure case study on nis}). For \MnIr, we additionally probe the magnetic anisotropy energy (Sec.~\ref{subsec:MAE}). 

To anticipate the main outcome as summarized in Sec.~\ref{sec:Conclusion}: the two SCDFT functionals are the only ones that recover the experimental noncollinear magnetic ground state of \NiS, but they do not yet uniformly reproduce other experimental observables with a higher accuracy than their locally collinear relatives. Overall, our results indicate that existing SCDFT functionals constitute a promising framework for studying noncollinear magnetism, while also underscoring the need for further validation across a wider range of experimental observables and magnetic systems.

\section{Methods}
\label{sec:Methods}

\subsection{Self-consistent field loops}
\label{subsec:ele min}

\begin{figure*}[t]
\centering
\resizebox{\textwidth}{!}{%
\begin{tikzpicture}

\begin{scope}
  \node[panelname] at (-7.6, 8.0) {(a) spin-DFT (locally collinear)};

  \node[box]     (ks)   at ( 4.4, -5.8)
    {\textbf{generalized Kohn--Sham equations}\\[2pt]
     $\big[-\tfrac12\nabla^2 + v_{\mathrm{ext}} + v_{\mathrm{H}}\big]
       \psi_{i\sigma} + \big(\hat v_{\mathrm{xc}}\psi_i\big)_{\sigma}
       = \varepsilon_i \psi_{i\sigma}$};
  \node[box]     (orb)  at (-5.4, -5.8) {\textbf{generalized Kohn--Sham spinors}\\[1pt] $\psi_{i\sigma}(\rv)$};
  \node[box]     (dens) at (-5.8, -2.2)
    {$\mathbf{2{\times}2}$ \textbf{densities}\\[2pt]
     $n_{\sg}=\sum_i f_i\, \psi_{i\sigma}^{*} \psi_{i\sigma'}$\\[1pt]
     \textcolor{mggacol}{$\tau_{\sg}=\tfrac12\sum_i f_i\,
        \nabla\psi_{i\sigma}^{*}\!\cdot\!\nabla\psi_{i\sigma'}$}};
  \node[diffbox] (proj) at (-5.8,  1.9)
    {\textcolor{sdftcol}{\textbf{project on local axis $\hat{m}(\rv)$}}\\[2pt]
     $n = \sum_{\sigma} n_{\sigma\sigma}$\\[1pt]
      $m^a = \sum_{\sg} \sigma^{a}_{\sigma'\sigma} n_{\sg}$\\[1pt]
     \textcolor{mggacol}{(likewise $\tau$, $\tau^a_m$)}\\[1pt]
     $n_{\uparrow\!/\!\downarrow}
        = \tfrac12\big(n \pm |\vec m|\big)$\\[1pt]
     \textcolor{mggacol}{$\tau_{\uparrow\!/\!\downarrow}
        = \tfrac12\big(\tau \pm \hat m\!\cdot\!\vec\tau_m\,\big)$}};
  \node[box]     (exc)  at (-1.1,  6.1)
    {\textbf{exchange-correlation functional}\\[1pt] locally collinear PZ, PW92, PBE, \textcolor{mggacol}{SCAN}\\[1pt]
     $E_{\mathrm{xc}}^{\mathrm{lc}}=\int \mathrm{d}\rv \, e_{\mathrm{xc}}^{\mathrm{lc}}(n_{\uparrow/\downarrow}, \nabla n_{\uparrow/\downarrow}, \textcolor{mggacol}{\tau_{\uparrow/\downarrow}})$};
  \node[box]     (var)  at ( 3.65,  2.2)
    {\textbf{exchange-correlation potentials}\\[2pt]
     $\big(\hat v_{\mathrm{xc}}\psi_i\big)_{\sigma}
       = \tfrac{\delta E_{\mathrm{xc}}}{\delta \psi_{i\sigma}^{*}}
       = \sum_{\sigma'} \Big[ v^{\mathrm{xc,mult}}_{\sg}\,\psi_{i\sigma'}
       - \textcolor{mggacol}{\nabla\!\cdot\!\big(\tfrac{\mu^{\mathrm{xc}}_{\sg}}{2}
           \nabla\psi_{i\sigma'}\big)}\Big]$\\[4pt]
     $v^{\mathrm{xc,mult}}_{\uparrow\!/\!\downarrow}
        = \tfrac{\partial e_{\mathrm{xc}}}{\partial n_{\uparrow\!/\!\downarrow}}
        - \nabla\!\cdot\!\tfrac{\partial e_{\mathrm{xc}}}
            {\partial \nabla n_{\uparrow\!/\!\downarrow}},
      \quad
      \textcolor{mggacol}{\mu^{\mathrm{xc}}_{\uparrow\!/\!\downarrow}
        = \tfrac{\partial e_{\mathrm{xc}}}{\partial \tau_{\uparrow\!/\!\downarrow}}}$};
  \node[diffbox] (ori)  at ( 4.8, -1.8)
    {\textcolor{sdftcol}{\textbf{orient along $\hat m(\rv)$}}\\[2pt]
     $v^{\mathrm{xc,mult}}_{\sg} = \bar v_{\mathrm{xc}}\,\delta_{\sg}
        + \Delta v_{\mathrm{xc}} \sum_a \sigma^{a}_{\sg}\, \hat m^{a}$\\[1pt]
        \textcolor{mggacol}{(likewise $\mu^{\mathrm{xc}}_{\sg}$)}\\[2pt]
     $\bar v_{\mathrm{xc}},\, \Delta v_{\mathrm{xc}}
        = \tfrac12 \big( v^{\mathrm{xc,mult}}_{\uparrow}
          \pm v^{\mathrm{xc,mult}}_{\downarrow} \big)$\\[2pt]
     $\Rightarrow\ \vec B_{\mathrm{xc}} \parallel \vec m$};

  \draw[flow] (ks.west)    -- node[above, font=\normalsize]{solve} (orb.east);
  \draw[flow] (orb.north)  to[bend left=20] (dens.south);
  \draw[flow] (dens.north) -- (proj.south);
  \draw[flow] (proj.north) to[bend left=22] (exc.west);
  \draw[flow] (exc.east)   to[bend left=22] ($(var.north west)!0.6!(var.north east)$);
  \draw[flow] ($(var.south west)!0.6!(var.south east)$)  -- (ori.north);
  \draw[flow] (ori.south)  to[bend left=20] (ks.north);

\end{scope}

\begin{scope}[shift={(19.6,0)}]
  \node[panelname] at (-7.6, 8.0) {(b) spin-current DFT};

  \node[box] (ksB)  at ( 4.8, -5.8)
    {\textbf{generalized Kohn--Sham equations}\\[2pt]
     $\big[-\tfrac12\nabla^2 + v_{\mathrm{ext}} + v_{\mathrm{H}}\big]
       \psi_{i\sigma} + \big(\hat v_{\mathrm{xc}}\psi_i\big)_{\sigma}
       = \varepsilon_i \psi_{i\sigma}$};
  \node[box] (orbB) at (-5.0, -5.8) {\textbf{generalized Kohn--Sham spinors}\\[1pt] $\psi_{i\sigma}(\rv)$};
  \node[box] (densB) at (-4.95,  0.0)
    {$\mathbf{2{\times}2}$ \textbf{densities}\\[2pt]
     $n_{\sg}=\sum_i f_i\, \psi_{i\sigma}^{*} \psi_{i\sigma'}$\\[1pt]
     $\tau_{\sg}=\tfrac12\sum_i f_i\,
        \nabla\psi_{i\sigma}^{*}\!\cdot\!\nabla\psi_{i\sigma'}$\\[2pt]
     \textcolor{scdftcol}{$\mathbf{j}_{\sg}=\tfrac{1}{2\mathrm{i}}\sum_i f_i
        \big[\psi_{i\sigma}^{*}\nabla\psi_{i\sigma'}
        - (\nabla\psi_{i\sigma}^{*})\,\psi_{i\sigma'}\big]$}};
  \node[box] (excB) at (-0.2,  6.1)
    {\textbf{exchange-correlation functional}\\[1pt] \textcolor{scdftcol}{NCMSCAN}, \textcolor{scdftcol}{LFNCBR89-NCCS}\\[1pt]
     $E_{\mathrm{xc}}^{\mathrm{SCDFT}}=\int \mathrm{d} \rv \, e_{\mathrm{xc}}^{\mathrm{SCDFT}}(n_{\sg}, \nabla n_{\sg}, \nabla^2 n_{\sg},\tau_{\sg},
        \textcolor{scdftcol}{\mathbf{j}_{\sg}})$};
  \node[box] (varB) at ( 4.9,  0.0)
    {\textbf{exchange-correlation potentials}\\[2pt]
     $\big(\hat v_{\mathrm{xc}}\psi_i\big)_{\sigma}
       = \tfrac{\delta E_{\mathrm{xc}}}{\delta \psi_{i\sigma}^{*}}
       = \sum_{\sigma'} \Big\{
        \big[v^{\mathrm{xc,mult}}_{\sg}
         + \textcolor{scdftcol}{\big(\nabla\!\cdot\!
             \tfrac{\mathbf{A}^{\mathrm{xc}}_{\sg}}{2\mathrm{i}}\big)}\big]\psi_{i\sigma'}$\\[2pt]
     $\quad- \nabla\!\cdot\!\big(\tfrac{\mu^{\mathrm{xc}}_{\sg}}{2} 
           \nabla\psi_{i\sigma'}\big)
       + \textcolor{scdftcol}{\tfrac{\mathbf{A}^{\mathrm{xc}}_{\sg}}{\mathrm{i}}
           \!\cdot\!\nabla\psi_{i\sigma'}} \Big\}$\\[4pt]
     $v^{\mathrm{xc,mult}}_{\sg}
        = \tfrac{\partial e_{\mathrm{xc}}}{\partial n_{\sigma'\sigma}}
        - \nabla\!\cdot\!\tfrac{\partial e_{\mathrm{xc}}}
            {\partial \nabla n_{\sigma'\sigma}}
        + \nabla^2 \tfrac{\partial e_{\mathrm{xc}}}
            {\partial \nabla^2 n_{\sigma'\sigma}}$\\[3pt]
     $\mu^{\mathrm{xc}}_{\sg} = \tfrac{\partial e_{\mathrm{xc}}}
        {\partial\tau_{\sigma'\sigma}}, \qquad
      \textcolor{scdftcol}{A^{\mathrm{xc}}_{\sg,\alpha}
        = \tfrac{\partial e_{\mathrm{xc}}}{\partial j_{\sigma'\sigma,\alpha}}}$};

  \draw[flow] (ksB.west)    -- node[above, font=\normalsize]{solve} (orbB.east);
  \draw[flow] (orbB.north)  to[bend left=20] ($(densB.south west)!0.25!(densB.south east)$);
  \draw[flow] ($(densB.north west)!0.25!(densB.north east)$) to[bend left=22] (excB.west);
  \draw[flow] (excB.east)   to[bend left=22] ($(varB.north west)!0.75!(varB.north east)$);
  \draw[flow] ($(varB.south west)!0.75!(varB.south east)$)  to[bend left=20] (ksB.north);

\end{scope}

\end{tikzpicture}%
}
\caption{Self-consistency cycle of (a) spin-DFT in the locally collinear
  approximation and (b) spin-current DFT. Starting from the generalized Kohn--Sham (GKS) spinors $\psi_{i\sigma}$, the $2{\times}2$ densities are computed, the exchange-correlation (XC) energy is evaluated, and its variation with respect to $\psi_{i\sigma}^{*}$ yields the XC potentials entering the GKS equations.
  Turquoise boxes mark the two steps exclusive to the locally collinear treatment: the projection of the densities onto the local spin-quantization axis $\hat m=\vec{m}/|\vec{m}|$ before the XC functional is evaluated, and the orientation of the resulting spin-diagonal potential along $\hat m$, which enforces zero XC torque. Purple denotes quantities specific to MGGA functionals (the kinetic-energy density $\tau_{\sg}$ and the associated potential $\mu^{\mathrm{xc}}$), and red the SCDFT-specific dependence on the paramagnetic spin-current density $\mathbf{j}_{\sg}$ with the resulting nonabelian XC vector potential $\mathbf{A}^{\mathrm{xc}}$. For brevity any details regarding the projector-augmented wave (PAW) method\cite{blochl1994projector,kresse1999ultrasoft,Huebsch2026:PRB:105101} or external vector potentials (representing external fields and/or spin-orbit coupling) are dropped. See Appendix~\ref{app:Notation} for explanations about the notation.}
\label{fig:scf-loop}
\end{figure*}
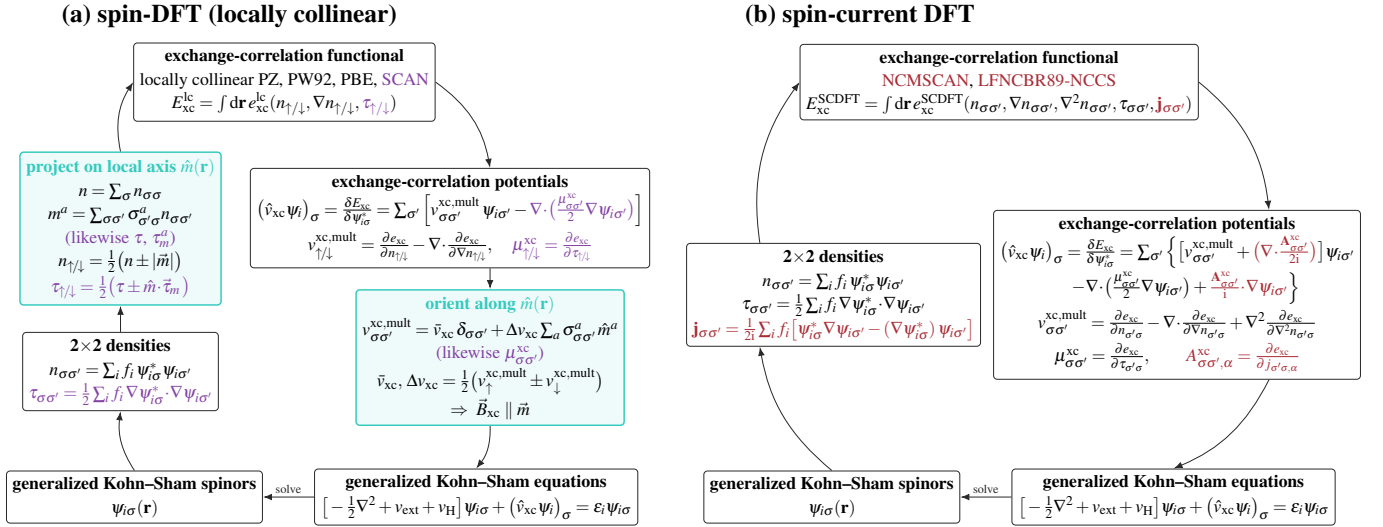

From the outset it is important to stress that while all calculations are noncollinear and include spin-orbit coupling self-consistently, the SCF loop underlying SCDFT is fundamentally distinct from that of locally collinear SDFT:

Figure~\ref{fig:scf-loop}(a) presents the locally collinear SDFT framework that was implemented over two decades ago in VASP.\cite{Hobbs2000:PRB:11556} It employs a trick proposed by Kübler to define a local spin-quantization axis which is found self-consistently by projecting the densities along the direction of the magnetization at each point in space.\cite{kubler1988density,kubler1988local} 
It corresponds to the conventional rotationally invariant construction for LSDA generalized to noncollinear magnetism. 
For lack of an equally simple prescription, locally collinear SDFT has also been applied to run locally collinear GGA and MGGA calculations.
The first instance of using a GGA was already reported in the implementation paper in Ref.~\onlinecite{Hobbs2000:PRB:11556}, although these extensions using GGA and MGGA do not capture the transverse spin gradients that characterize a truly noncollinear state---for an analysis of the exchange energy, see Ref.~\onlinecite{EichPhysRevB.88.245102}. However, this shortcoming could not be quantified, while the benefit of using GGA and MGGA for quantities such as band gap, lattice constants, etc.\ was clearly tangible.\cite{Perdew1992:PRB:6671,perdew1996generalized,sun2015strongly}

We also should highlight that here we work within the generalized Kohn--Sham\cite{Seidl1996:PRB:3764} (GKS) formulation, where the energy is minimized with respect to variations of the two-component spinors, in terms of which the densities and currents are expressed.\cite{Desmarais2019:JPCL:3580,Desmarais2024:PRM:13802} 
Alternatively, differentiation with respect to the basic densities of spin-orbital functionals can be performed in the fashion of the optimized effective potential method,
see, for example, Refs.~\onlinecite{Pittalis2006,Goerling2006,HeatonBurgess2007,Sharma2007a,Sharma2007b,Goerling2018}.

Only recently, the noncollinear SCF loop underlying SCDFT\cite{vignale1988current} shown in Fig.~\ref{fig:scf-loop}(b) was implemented in VASP; see Ref.~\onlinecite{Huebsch2026:PRB:105101}. Compared to the SDFT version shown in Fig.~\ref{fig:scf-loop}(a), it entails additional terms in the variation of the XC energy with respect to the GKS spinor. 
In particular, the approach in Fig.~\ref{fig:scf-loop}(b) does not rely on the projection onto a local spin-quantization axis as employed in Fig.~\ref{fig:scf-loop}(a).

Noncollinear SCDFT functionals do not just depend on additional quantities on top of LSDA, like higher-order derivatives of the charge density, the full noncollinear magnetization, the kinetic-energy density, and the paramagnetic spin-current density; they also treat the noncollinearity of the system on a fundamentally different footing: the functional depends on $2\times 2$ densities and yields $2\times 2$ functional derivatives, while the XC torques it generates are exactly constrained by the U(1)$\times$SU(2) invariance.\cite{Desmarais2026}

\subsection{Exchange-correlation approximations}
\label{subsec:XC functionals}

In the locally collinear construction, the semi-local XC energy has the form 
\begin{equation}
 E_\mathrm{xc}^\mathrm{lc}
 = \int \diff^3r\,
 e_\mathrm{xc}^\mathrm{lc}
 \!\left(n_{\uparrow/\downarrow},\nabla n_{\uparrow/\downarrow},
 \tau_{\uparrow/\downarrow}\right).
 \label{eq:lc-dependencies}
\end{equation}
In VASP, the spin-resolved quantities are obtained by projecting the spin-density and kinetic-energy-density matrices onto the direction of the magnetization at that point in space:

\begin{subequations}
\begin{equation}
\label{eq:n}
 n_{\uparrow/\downarrow}=\frac{1}{2}\left(n\pm|\vec m|\right),
\end{equation}
\begin{equation}
\label{eq:tau}
 \tau_{\uparrow/\downarrow}=\frac{1}{2}
 \left(\tau\pm\hat m\!\cdot\!\vec\tau\right),
\end{equation}
\end{subequations}
where $\hat m=\vec m/|\vec m|$.

Although the generated heuristic extensions represent distinct functionals, we loosely reuse the bare names PZ, PW92, PBE, and SCAN from the original collinear functionals. It is encouraged by the fact that the formulation indeed reproduces collinear results in the collinear limit. The density projection above gives the conventional rotationally invariant generalized LSDA construction for PZ and PW92.

The requirement of exact constraints to guide the construction of noncollinear density functional approximations motivates the broader variational framework of SCDFT. The SCDFT approximations considered here have the semi-local form

\begin{equation}
 E_\mathrm{xc}^\mathrm{SCDFT}
 = \int \diff^3r\,
 e_\mathrm{xc}^\mathrm{SCDFT}
 \!\left(n_{\sigma\sigma'},\nabla n_{\sigma\sigma'},
 \nabla^2 n_{\sigma\sigma'},\tau_{\sigma\sigma'},
 \mathbf j_{\sigma\sigma'}\right)
 \label{eq:scdft-dependencies}
\end{equation}
and depend explicitly on the paramagnetic current matrix $\mathbf j_{\sigma\sigma'}$. Figure~\ref{fig:scf-loop}(b) defines the matrix quantities and shows how the different dependencies in Eqs.~(\ref{eq:lc-dependencies}) and (\ref{eq:scdft-dependencies}) enter the two SCF cycles.

The basic variables for SCDFT are $n$, $m^a$, the paramagnetic charge current $j_\nu$, and the paramagnetic spin current $J^a_\nu$. 
\cite{vignale1988current,bencheikh2003spin}
Their conjugate XC potentials are, respectively, the scalar XC potential $v_{\mathrm{xc}}=\delta E_{\mathrm{xc}}/\delta n$, the XC magnetic field $B^a_{\mathrm{xc}}=\delta E_{\mathrm{xc}}/\delta m^a$, the XC Abelian vector potential $A_{\mathrm{xc},\nu}=\delta E_{\mathrm{xc}}/\delta j_\nu$, and the non-Abelian XC vector potential $A^a_{\mathrm{xc},\nu}=\delta E_{\mathrm{xc}}/\delta J^a_\nu$. The invariance of the XC energy under local U(1)$\times$SU(2) gauge transformations constrains the XC potentials and provides an exact design condition for SCDFT approximations. After a deeper analysis one can show that SCDFT permits nonzero internal XC torques that help determine the noncollinear ground state without appearing as spurious terms in adiabatic spin dynamics.\cite{Desmarais2026}

In a nutshell, all calculations in this work are based on six semi-local XC approximations. This includes the locally collinear extensions of PZ and PW92 at the LSDA level, PBE at the GGA level, and SCAN at the MGGA level. The locally collinear prescriptions used for these four approximations are specified by Eqs.~(\ref{eq:lc-dependencies}), (\ref{eq:n}), and (\ref{eq:tau}) and in Fig.~\ref{fig:scf-loop}(a). The two remaining functionals, LFNCBR89-NCCS and NCMSCAN, are SCDFT approximations, corresponding to Eq.~(\ref{eq:scdft-dependencies}) and Fig.~\ref{fig:scf-loop}(b). Details about the functionals are provided in Secs.~\ref{subsec:Collinear} and \ref{subsec:Noncollinear}.

\subsubsection{Locally collinear functionals}
\label{subsec:Collinear}

The PZ\cite{perdew1981self} and PW92\cite{Perdew1992:PRB:13244} LSDA functionals combine Slater exchange \cite{dirac1930note} with parameterizations of Monte Carlo correlation data for the homogeneous electron gas.\cite{ceperley1980ground}
For historical reasons, PZ is the most used LSDA within the VASP community,\cite{kresse1999ultrasoft} while PW92 is the LSDA component of numerous GGA and MGGA functionals.
Notably, the spin dependence of PW92 is built on the RPA correlation spin stiffness of the uniform electron gas.
We mention that PZ and PW92 are implemented in VASP with a Slater exchange that includes a relativistic correction,\cite{MacDonald1979:JPCSSP:2977}
which is not the case in the Libxc implementations,\cite{Lehtola2018:S:1} for instance. Here we use PZ with relativistic correction and PW92 via Libxc without relativistic correction. 
Also note that PZ is known to be numerically less stable than PW92, as discussed in Ref.~\onlinecite{Lehtola2022:JCP:174114}, a behavior that we also observed in our calculations (see Sec.~\ref{subsec:Computational cost}).

PBE\cite{perdew1996generalized} is the most popular GGA functional in the solid-state community.\cite{Jain2013:APLM:011002} Constructed to be analytically simpler than its predecessor PW91,\cite{Perdew1992:PRB:6671} PBE is much more useful than the LSDA for geometry optimization and energetics.\cite{perdew1996generalized} The GGA set of PAW potentials was constructed based on PBE so that it evolved to be the default GGA in VASP. Many GGA and MGGA functionals that have been subsequently proposed are based on PBE.\cite{tran2026semi} 

SCAN was constructed such that it obeys numerous constraints within DFT at the level of a collinear MGGA.\cite{sun2015strongly} SCAN is prone to be numerically less stable in VASP as it involves step functions and thus often requires a dense integration grid, i.e., within the PAW method a higher cutoff energy for plane-wave basis sets, compared to other functionals.\cite{Yang2016:PRB:205205,Bartok2019:JCP:161101} This problem has motivated the construction of \emph{regularized} versions of SCAN.\cite{Bartok2019:JCP:161101,Furness2020:JPCL:8208,Furness2022:JCP:34109} SCAN and its variant R2SCAN\cite{Furness2020:JPCL:8208} are frequently used for calculations on materials.\cite{Horton2025:NM:1522}

\subsubsection{Noncollinear SCDFT functionals}
\label{subsec:Noncollinear}

After the seminal works from Vignale and Rasolt\cite{Vignale1987:PRL:2360,vignale1988current} on the SCDFT formalism and the studies from Dobson\cite{Dobson1991:JCP:4328,Dobson1992:JPCM:1992,Dobson1993} on the Fermi hole curvature and its relation to the charge-current density, early attempts to include the paramagnetic charge-current or collinear spin-current densities into XC functionals were made by Becke and co-workers,\cite{Becke1996:CJC:995,Becke2002:JCP:6935,Johnson2007:JCP:184104} Maximoff, Ernzerhof, and Scuseria,\cite{Maximoff2004:JCP:2105} Tao and Perdew,\cite{Tao2005:PRB:205107,Tao2005:PRL:196403} Pittalis \textit{et al.},\cite{Pittalis2007:PRB:235314,PittalisPhysRevA.80.032515,Pittalis2007:JCP:124103} and Bates and Furche,\cite{bates2012harnessing} the latter in the time-dependent context. Note that these functionals were not developed in a noncollinear context. There has been renewed interest in the development of such current-density functionals in recent years, and the most recent works --- now mostly in a noncollinear context, yet without satisfying U(1)$\times$SU(2) invariance --- include those from 
Tancogne-Dejean, Rubio, and Ullrich,\cite{tancogne2023constructing,tancogne2023constructing_erratum} and Franzke, Pausch, and Holzer.\cite{Franzke2025:JCP:084104}

NCMSCAN and LFNCBR89-NCCS, the two SCDFT approximations introduced in Sec.~\ref{sec:Introduction}, which are of the form given by Eq.~(\ref{eq:scdft-dependencies}), were designed to satisfy the local U(1)$\times$SU(2) gauge invariance. They are based on existing locally collinear functionals, and a short summary of the changes made in their parent functionals to derive them is provided below. The detailed derivations can be found in Refs.~\onlinecite{Pittalis2017:PRB:035141,desmarais2024electron,Desmarais2025:PRL:106402,tancogne2023constructing,tancogne2023constructing_erratum,Huebsch2026:PRB:105101}, where discussions on various aspects --- including the derivation of the extended XC spin-torques in SCDFT~\cite{Desmarais2026} --- can also be found. The notation used in the equations below is explained in Appendix~\ref{app:Notation}.

NCMSCAN was derived from SCAN\cite{sun2015strongly} by replacing the spin-polarization
\begin{equation}
\zeta=\frac{n_{\uparrow}-n_{\downarrow}}{n}
\end{equation}
and iso-orbital indicator (connected to the electron localization function\cite{becke1990simple})
\begin{equation}
\alpha=\frac{\tau-\tau_{\rm W}}{\tau_{\rm unif}},
\end{equation}
where $\tau_{\rm unif}=(3/10)(3\pi^2)^{2/3}n^{5/3}$ and $\tau_{\rm W}=\left\vert\nabla n\right\vert^2/(8n)$,
by modified noncollinear versions:
\begin{equation}
\zeta^{\rm nc} = \frac{\vert \vec{m} \vert}{n},
\end{equation}
\begin{equation}
\label{eq:alphanc}
\alpha^{\rm nc} = \frac{ n\left(\tilde{\tau}^{\rm nc}- \tau_{\rm W}\right)}{  n_\uparrow \tau_{\rm unif}^{\uparrow} +   n_\downarrow\tau_{\rm unif}^{\downarrow}  },
\end{equation}
respectively, where\cite{Pittalis2017:PRB:035141} 

\begin{equation}
\label{eq:tTAU}
 \tilde{\tau}^{\rm nc} =   
 \tau  +   \frac{\vec{m} \cdot \vec{\tau}_{m}}{n}
 +  \frac{|\mathbf{\nabla} \vec{m}|^2}{8 n}  - \frac{|\mathbf{j}|^2}{ 2n} - \frac{ |\vec{ \mathbf{J}}|^2}{2n}
\end{equation} 
and
\begin{equation}
\tau_{\rm unif}^{\uparrow/\downarrow} = \frac{3}{10}(3\pi^2)^{2/3}(2n_{\uparrow/\downarrow})^{5/3}.
\end{equation}
Note that the denominator in Eq.~(\ref{eq:alphanc}) can be expressed entirely by $n$ and $|\vec{m}|$ and requires no projection in the sense of Eq.~(\ref{eq:tau}).
Desmarais \textit{et al.}\cite{Desmarais2025:PRL:106402} showed that NCMSCAN leads to smaller magnetic moments compared to SCAN, which improves the agreement with experiment for itinerant metals like Fe, Co, and Ni, and that no spurious moment in non-magnetic systems like bulk V and Pd or the molecules C$_2$ and Cr$_2$ is obtained.

LFNCBR89-NCCS is based on the BR89\cite{becke1989exchange} and CS\cite{colle1975approximate} functionals for the exchange and correlation parts, respectively. Converting the BR89 expression into a noncollinear one was done by replacing the density $n$ and exchange-hole curvature for closed-shell systems
\begin{equation}
\label{eq:Qx}
    Q_\text{x} = \frac{1}{6}\left[\frac{1}{2}\nabla^2 n -2\gamma \left(\tau - \tau_{\rm W}\right)\right]
\end{equation}
by
\begin{equation}
 \label{eq:ntop}
    n_\text{top} = \frac{n}{2} \left(1+\frac{|\vec{m}|^2}{n^2}\right),
\end{equation}
\begin{equation}
    \label{eq:Q tilde}
    Q_{\text{x}}^{\rm nc}=\frac{1}{6} \left[ -8 \tau_\text{W}^\text{nc}  -4 \gamma^{\rm nc} \left( 
    \bar \tau^\text{nc} - \tau_\text{W}^\text{nc}\right) \right],
\end{equation}
respectively, where
\begin{equation}
\label{eq:taunm}
    \bar{\tau}^\text{nc} = \frac{1}{2}\left(\tau + \frac{\vec{m} \cdot \vec \tau_{m}}{n}
    - \frac{|\mathbf{j}|^2}{ 2n} - \frac{ |\vec{ \mathbf{J}}|^2}{2n}\right)
\end{equation}
and 
\begin{align}
    \label{eq:tauwncl}
    \tau_\mathrm{W}^\mathrm{nc} = \frac{\tau_\mathrm{W}}{2}\left(1+\frac{|\mathbf{\nabla} \vec{m}|^2}{\left\vert\mathbf{\nabla} n\right\vert^2}\right).
\end{align}
Two points should be made about Eq.~(\ref{eq:Q tilde}): (i) It is a Laplacian-free expression that was obtained by integration by parts of the exchange energy, and (ii) satisfaction of the U(1)$\times$SU(2) gauge invariance requires $\gamma^{\rm nc}=1$ (the value used in Ref.~\onlinecite{Huebsch2026:PRB:105101} and in the present work). For the collinear expression in Eq.~(\ref{eq:Qx}), Becke and Roussel argued for $\gamma=0.8$ in Ref.~\onlinecite{becke1989exchange}. Note that the two definitions of the noncollinear kinetic-energy density used in NCMSCAN and LFNCBR89-NCCS are related by
\begin{equation}
    \bar{\tau}^\text{nc} = \frac{1}{2}\left(\tilde\tau^{\rm nc} + \tau_{\rm W} -2\tau_{\rm W}^{\rm nc}\right),
\end{equation}
i.e., they differ by the term $|\mathbf{\nabla} \vec{m}|^2/(8 n)$.
The factor $1/2$ is unimportant and just comes from the different conventions of $\tilde\tau^{\rm nc}$ recovering $\tau$, $\bar\tau^{\rm nc}$ recovering $\tau/2$, and $\tau_{\rm W}^{\rm nc}$ recovering $\tau_{\rm W}/2$ in the unpolarized case, cf.~Eqs.~(\ref{eq:tTAU}), (\ref{eq:taunm}), and (\ref{eq:tauwncl}).

The CS formula for the correlation energy depends on the two-electron reduced density matrix $\rho_2$. By using the expression for $\rho_2$ in terms of the 
one-electron reduced density matrix $\rho_1$, given by (the unpolarized form is shown)
\begin{equation}
    \rho_{2}(\mathbf{r}_1,\mathbf{r}_2;\mathbf{r}_1,\mathbf{r}_2)= n(\mathbf{r}_1)n(\mathbf{r}_2)-\frac{1}{2}\rho_{1}(\mathbf{r}_1,\mathbf{r}_2)\rho_{1}(\mathbf{r}_2,\mathbf{r}_1),
\end{equation}
Lee, Yang, and Parr\cite{Lee1988lyp} obtained an MGGA functional that depends on $\tau$ and $\nabla^2 n$
(the dependency on $\nabla^2 n$ was later eliminated,\cite{Miehlich1989:CPL:200} leading to the famous LYP functional).
For a noncollinear version, Tancogne-Dejean, Rubio, and Ullrich\cite{tancogne2023constructing,tancogne2023constructing_erratum} used
\begin{equation}
\label{eq:CS rho2HF ncl}
    \rho_{2}^{\rm nc}(\mathbf{r}_1,\mathbf{r}_2;\mathbf{r}_1,\mathbf{r}_2)= n(\mathbf{r}_1)n(\mathbf{r}_2)-\sum_{\sigma\sigma'}\rho^{\sigma\sigma'}_{1}(\mathbf{r}_1,\mathbf{r}_2)\rho^{\sigma'\sigma}_{1}(\mathbf{r}_2,\mathbf{r}_1),
\end{equation}
and combined it with a noncollinear version of the exchange BR89 functional (different from the one used in LFNCBR89-NCCS).

\subsection{Computational details}

\begin{table*}[t]
  \centering
  \caption{Plane-wave cutoff energies (eV) and $\mathbf{k}$-mesh dimensions for each functional and material.}
  \label{tab:encut}
  \begin{ruledtabular}
  \begin{tabular}{lccccccc}
       & PZ & PW92 & PBE & SCAN  & NCMSCAN & LFNCBR89-NCCS & $\mathbf{k}$ mesh\\
    \midrule
    \MnIr  & 700 & 550 & 600 & 800  &  750 &  800 & $7\times7\times7$\\
    \MnGe  & 750 & 750 & 800 & 850  &  850 &  800 & $5\times5\times7$\\
    \NiS   & 800 & 750 & 700 & 1250  & 1250 & 1100 & $7\times7\times7$\\
    \YMO   & 700 & 700 & 750 & 750 &  700 &  700 & $4\times4\times2$\\
  \end{tabular}
  \end{ruledtabular}
\end{table*}

We employ the projector-augmented wave (PAW) method.\cite{blochl1994projector,kresse1999ultrasoft} In other words, VASP uses basis functions for the GKS spinors that comprise plane waves and a local on-center part to describe both the periodicity of the bulk systems and the nodal features characteristic of electrons with angular momentum quantum number $l\ge2$. Some of the electrons are treated as core electrons and are not relaxed during the SCF cycle. The PAW potentials are the standard pseudopotentials distributed with VASP: Mn\_pv, Ni\_sv\_GW, Ir, Ge\_d, S, Y\_sv, and O.

Overall, the computational parameters are converged such that increasing the plane-wave cutoff or the $\mathbf{k}$-mesh density changes the total energy by less than $1$\,meV/atom.
The plane-wave cutoff energies are determined from per-functional convergence studies on each material and are summarized in Table~\ref{tab:encut}. The Brillouin zone is sampled with $\Gamma$-centered Monkhorst--Pack\cite{monkhorst1976special,pack1977special} $\mathbf{k}$-meshes of equal dimensions for all functionals, also listed in Table~\ref{tab:encut}. The stopping criterion for the electronic minimization is that the sum over the change in band energies is less than $10^{-8}$\,eV.

\subsection{Structure parameters and volume relaxation}
\label{subsec:Volume relaxation}

\begin{figure*}[t]
  \centering
  \includegraphics[width=0.95\textwidth]{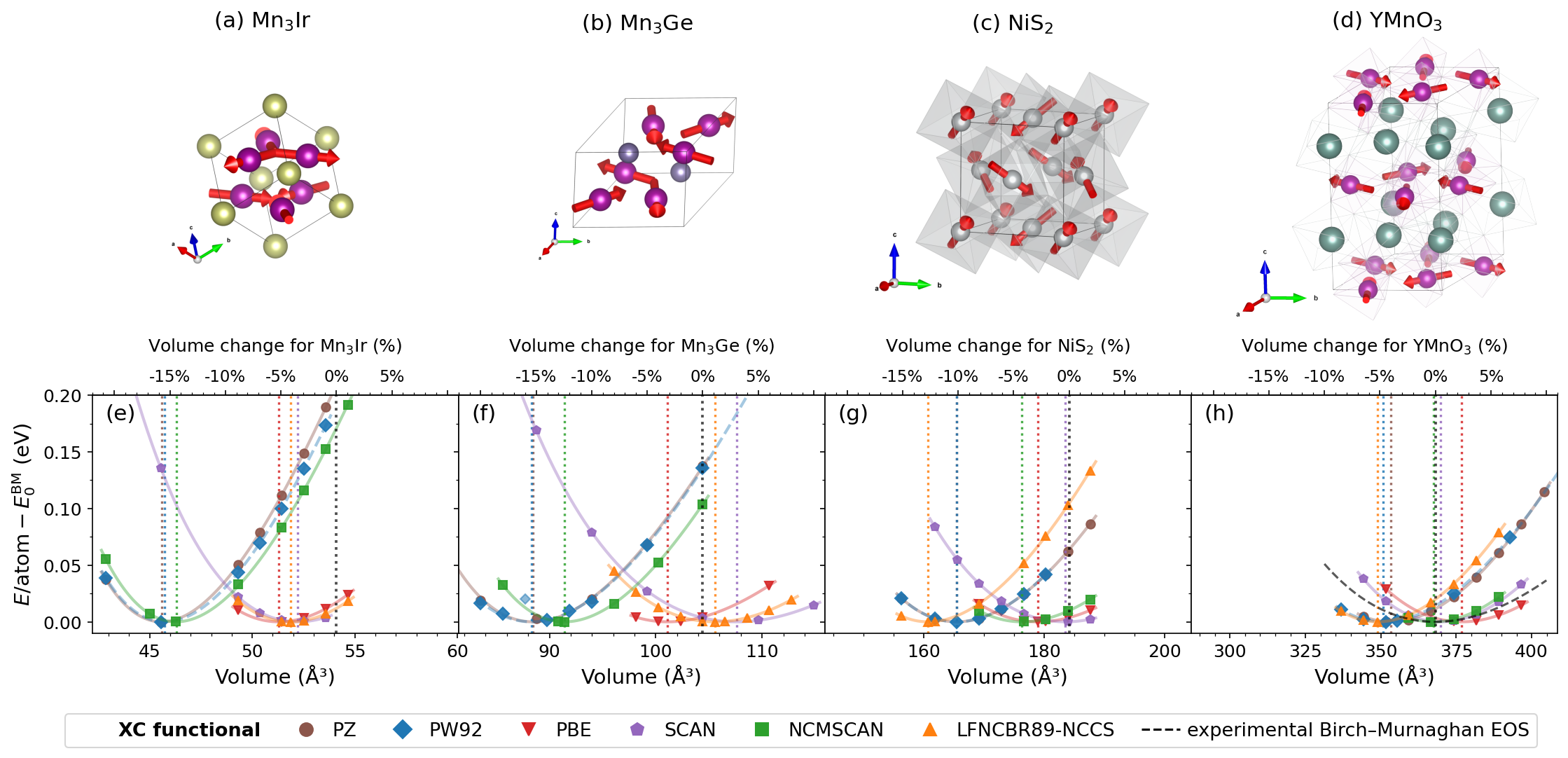}
  \caption{Structure, magnetic order, and equations of state of the four benchmark materials.
  (a)-(d)~Conventional cells of \MnIr, \MnGe, \NiS, and \YMO\ with the
  experimentally determined magnetic configurations (red arrows) of
  MAGNDATA entries, cf.~Table~\ref{tab:exp lattice}.
  (e)-(h)~Energy per atom, relative to the minimum $E_0^\mathrm{BM}$ of
  the respective Birch--Murnaghan (BM) fit, as a function of the
  cell volume for the six XC functionals.  The cells are
  scaled isotropically, with the internal coordinates and (for the
  hexagonal systems) the $c/a$ ratio fixed to the experimental
  structure.  Symbols are computed total energies; the curves are the BM fits,
  drawn dashed for PW92 where it nearly coincides with PZ.  Colored
  dotted vertical lines mark each functional's equilibrium volume
  $V_\mathrm{BM}$.  The black dotted line is the experimental reference
  volume $V_0$ of Table~\ref{tab:exp lattice} and the top axes give the volume change relative to
  it.  For \YMO\ the black dashed curve is the experimental EOS
  constructed from the measured bulk modulus of $112$\,GPa~\cite{gao2011high} (with $B'=4$) centered at $V_0$.}
  \label{fig:eos}
\end{figure*}

\begin{table*}[t]
  \centering
  \caption{Experimental reference data for the benchmark materials: MAGNDATA entry id,
    parent (nonmagnetic) space group, magnetic space group, corresponding BNS
    setting, experimentally measured on-site magnetic moment $m_\mathrm{expt}$ in $\muB$, lattice constants $a$ and $c$ in \AA, and corresponding
    conventional-cell volume $V$ in \AA$^3$ as used as initial parameters for the
    calculation, the antiferromagnetic transition temperature $T_N$ and the temperature at which the magnetic order was determined $T$ in K, lattice constants $a_0$ and $c_0$ in \AA, and corresponding
    conventional-cell volume $V_0$ in \AA$^3$ for $T\to0$ as discussed in the main text, bulk modulus $B$ in GPa, and electronic band gap $E_{g,\mathrm{expt}}$ in eV.}
  \label{tab:exp lattice}
  \begin{ruledtabular}
  \begin{tabular}{lcccc}
                & \MnIr        & \MnGe      & \NiS        & \YMO       \\
    \colrule
    MAGNDATA entry    & 0.108\cite{tomeno1999magnetic}        & 0.377\cite{soh2020ground}       & 0.150\cite{yano2016magnetic}       & 0.6\cite{munoz2000magnetic}       \\ \\
    Parent space group   & $Pm\bar{3}m$  & $P6_3/mmc$ & $Pa\bar{3}$ & $P6_3cm$   \\
                & (221)        & (194)      & (205)       & (185)      \\ \\
    Magnetic space group  & $R\bar{3}m'$  & $Cm'cm'$\cite{tomiyoshi1983triangular,nagamiya1982triangular,chen2020antichiral}   & $Pa\bar{3}$ & $P6_3cm$   \\
    (BNS)             & (166.101)        & (63.464)   & (205.33) & (185.197)  \\ 
    $m_\mathrm{expt}$ ($\muB$) & ---\footnotemark[6] & 2.65(2)\footnotemark[5] & 0.98(6)\cite{yano2016magnetic} &  2.90(2)\cite{munoz2000magnetic} \\ \\
    $a$ (\AA)   & 3.770              & 5.302      & 5.687  & 6.155      \\
    $c$ (\AA)   & ---                   & 4.289      & --- & 11.403     \\
    $V$ (\AA$^3$) & 53.58           & 104.42     & 183.93    & 374.14     \\ \\
    $T_N$ (K)      & 960(10)\cite{tomeno1999magnetic}         &  380\cite{kiyohara2016giant,nayak2016large}        & 39.2\cite{kikuchi1978spin,thio1995surface}          & 70\cite{munoz2000magnetic} \\
    $T$ (K) of magnetic & 300  & 2 & 2  & 1.7\\ 
    structure determination  & \\ \\
    For $T \to 0$\,K: & & & \\
    $a_{0}$ (\AA)   & 3.782\footnotemark[1]
                         & ---\footnotemark[2]
                         & 5.689\footnotemark[3]
                         & 6.115\footnotemark[4] \\
    $c_{0}$ (\AA)   & ---
                         & ---\footnotemark[2]
                         & ---
                         & 11.367\cite{munoz2000magnetic} \\
    $V_{0}$ (\AA$^3$) & 54.10 & 104.42\footnotemark[2] & 184.12 & 368.1 \\
    $B$  (GPa)    & ---    & ---   & ---        & 112(8)\cite{gao2011high}  \\ \\
    $E_{g,\mathrm{expt}}$ (eV) & --- & --- & 0.3\cite{sato1984reflectivity} & 1.5\cite{kalashnikova2003electronic} \\
  \end{tabular}
  \end{ruledtabular}
  \footnotetext[1]{Quadratic $T\to 0$ extrapolation of the 292--700\,K data
    in Fig.~7 of Ref.~\onlinecite{tomeno1999magnetic}.}
  \footnotetext[2]{Using the MAGNDATA entry for lack of a better alternative. $a'_0=5.353$~\AA, $c'_0=4.343$~\AA, and $V'_0=107.77$~\AA$^3$ are obtained from the Debye--Gr\"{u}neisen fit ($\theta_D=350$\,K) for the Mn-rich composition Mn$_{3.2}$Ge.\cite{sukhanov2018gradual}}
  \footnotetext[3]{Reported in Ref.~\onlinecite{yano2016magnetic} for 2\,K at ambient pressure (same experiment as MAGNDATA entry 0.150, which lists $a=5.687$\,\AA).}
  \footnotetext[4]{Low-temperature plateau in Fig.~12 of
    Ref.~\onlinecite{munoz2000magnetic}.}
  \footnotetext[5]{Soh \emph{et al.}\cite{soh2020ground} spherical neutron polarimetry at zero field. More precise than unpolarized powder neutron diffraction: 2.52\,$\muB$\cite{sukhanov2018gradual} measured earlier.}
  \footnotetext[6]{Hard to determine experimentally due to neutron absorption of Ir. See discussion in Appendix~\ref{app:Structural parameters}.}
\end{table*}

The crystal structures are relaxed for each XC functional. The starting points are the experimental measurements listed in MAGNDATA:\cite{gallego2016magndata} \MnIr\ entry 0.108,\cite{tomeno1999magnetic} \MnGe\ entry 0.377,\cite{soh2020ground} \NiS\ entry 0.150,\cite{yano2016magnetic} and \YMO\ entry 0.6.\cite{munoz2000magnetic} The structures are shown in Figs.~\ref{fig:eos}(a)--(d) including the experimentally determined magnetic configurations on the Mn and Ni sites. The experimental lattice parameters are summarized in Table~\ref{tab:exp lattice}. The Birch--Murnaghan (BM) equation-of-state (EOS)\cite{birch1947finite,murnaghan1944compressibility} curve corresponding to the experimental lattice constants and bulk modulus is shown as a black dashed line in Fig.~\ref{fig:eos}(h). The choice of the $T\to0$ reference parameters is discussed in Appendix~\ref{app:Structural parameters}.

In order to determine the relaxed volume for each XC functional, we perform static calculations at different fixed volumes by scaling all three lattice vectors with the initial magnetic configuration set to the experimental one and keeping the ionic positions fixed to their fractional coordinates. For the hexagonal systems, we verified for PW92 and PBE that also relaxing the ratio $c/a$ changes the relaxed volume by $\ll 1\%$, hence on a scale negligible for our analysis. To obtain accurate total energies, we first converge with the conjugate gradient electronic minimizer with Gaussian smearing and then restart from the converged wave functions using the improved tetrahedron method with Blöchl corrections.\cite{blochl1994improved} Finally, we fit the BM EOS using the atomic simulation environment (ASE)\cite{hjorth2017atomic} to extract the relaxed volume and bulk modulus. The per-functional results are discussed in Sec.~\ref{subsec:Structural properties}. 

\subsection{Magnetic order}
\label{subsec:mag order}

For convergence studies and volume relaxation, we use the experimental magnetic configuration as initialization for the magnetic moments as listed in MAGNDATA, cf.~Figs.~\ref{fig:eos}(a)--(d). 
For the analysis of the energy landscape with respect to magnetic configurations, we employ the cluster-multipole expansion\cite{suzuki2017cluster,suzuki2019multipole} and take a closer look at particularly interesting linear combinations of configurations with the same irreducible representation and multipole expansion order\cite{huebsch2021benchmark} for \MnIr\ and \NiS. The magnetic configurations are shown in Figs.~\ref{fig:landscape}(b) and (d) and are also tabulated in Appendix~\ref{app:Magnetic configurations}.

Calculations at different magnetic configurations are performed using only the appropriate initialization of the magnetic moments. Unless stated otherwise, no constraints are applied to the magnetic configuration. For constrained calculations, we constrain the direction and sign of the magnetic moment\cite{ma2015constrained} based on an energy penalty added to the (SC)DFT energy.\footnote{M. Marsman, unpublished.} This entails multiple total energy calculations at different penalty couplings $\lambda$ to obtain a total energy $E(\lambda)$ that includes the penalty term. One finds that $E(\lambda)-E_{\mathrm{(SC)DFT}} \propto \frac{1}{\lambda}$ and, hence, to extract the (SC)DFT total energy excluding the energy penalty, the function $E(\lambda)= \frac{\alpha}{\lambda} + E_{\mathrm{(SC)DFT}}$ is fitted via the fitting parameters $\alpha$ and $E_{\mathrm{(SC)DFT}}$. The magnetic energy landscape is presented in Secs.~\ref{subsec:Magnetic energy landscape} and \ref{subsec:MAE}.

\subsection{Computational cost}
\label{subsec:Computational cost}

\begin{table}[t]
  \centering
  \caption{Relative cost of the functionals. Column (A)
    gives the median wall time per electronic step in a
    controlled benchmark (\MnIr{} in the experimental cell, 10
    repetitions of 10 \texttt{ALGO}$=$\texttt{Damped} steps
    restarted from a 5-step Davidson \texttt{WAVECAR}, no
    wavefunction/charge IO, single exclusive node, 32 MPI ranks)
    at a common cutoff of 800\,eV, relative to the PW92
    repetition of the same time block, with the standard
    deviation over the 10 repetitions in parentheses.  Column (B) 
    gives the median for the time to solution over all materials
    and two volume points (the experimental and the BM minimum), of the total CPU core time ($t\,N_\mathrm{MPI}$) of the EOS workflow runs, relative to the PW92 run of the same material and volume, with the median absolute deviation in parentheses in units of the last digit.}
  \label{tab:cost summary}
  \begin{ruledtabular}
  \begin{tabular}{lcc}
       & (A)  & (B)\\
       & cost per SCF step & time to solution\\
    \midrule
    PZ & 0.931(5) & 1.4(4)\\
    PW92 & 1 & 1\\
    PBE & 1.005(5) & 1.5(7)\\
    SCAN & 4.62(2) & 9(6)\\
    NCMSCAN & 6.96(15) & 7(3)\\
    LFNCBR89-NCCS & 6.92(3) & 5(2)\\
  \end{tabular}
  \end{ruledtabular}
\end{table}

Regarding the computational cost, we want to present two perspectives:

\begin{itemize}
    \item[(A)] The cost per SCF step in a \emph{controlled benchmark}. It entails artificially setting as many parameters as possible to the same value across the benchmark, e.g., the plane-wave cutoff, the volume, the number of SCF steps, and choosing the algorithm for its reproducible computational effort.
    \item[(B)] The time to solution in a realistic setting. Here, the parameters are set according to convergence studies, and the SCF loop runs until the stopping criterion, i.e., the sum over the change in band energies, is satisfied. Moreover, the algorithm is chosen for its flexibility in finding the fastest and most reliable path to a minimum: the optimizer may, e.g., perform steepest-descent steps in between conjugate-gradient steps, or refine the line search. We probe this \emph{on-road consumption} simply by evaluating a subset of the calculations performed for the EOS workflow.
\end{itemize}

More details on the procedure for (A) and (B) are discussed in Appendix~\ref{app:Cost estimation}. Generally, all six XC approximations are semi-local, which keeps their computational cost well below that of hybrid functionals\cite{Becke1993:JCP:1372,Becke1993:JCP:5648} and many-body approaches such as $GW$.\cite{Hedin1965:PR:796}
Table~\ref{tab:cost summary} presents the resulting cost factors normalized to the run time of PW92. 

For perspective (A), we first note a difference between PZ and PW92. That is because PW92 has a small overhead due to interfacing with Libxc. The increasing cost per SCF step of going from LSDA to GGA, MGGA and SCDFT is dominated by the extra terms in the Hamiltonian application (cf.~XC potentials box in Fig.~\ref{fig:scf-loop}) and the associated FFTs. SCDFT needs exactly twice as many FFTs as MGGA. Also among the two SCDFT functionals we see a tiny difference, which vanishes within the uncertainty. It originates in the fact that LFNCBR89-NCCS requires some extra time to compute the Laplacian, while NCMSCAN has a heavier kinetic-energy-density construction.

Perspective (B) is more relevant in practice and the handle where functional design can pay off. The dominating effect is how many SCF steps are necessary to converge. To a lesser degree the functional-dependent choice of the plane-wave cutoff energy, the stability of the SDFT vs SCDFT loops, and the size of the unit cell at the EOS minimum play a role. To avoid outliers dominating the picture, column (B) reports the median and, in parentheses, the median absolute deviation of the cost factor for a full workflow normalized to the run time of PW92. We see that the deviation is much larger, making an estimate of the computational cost for a project quite hard. For instance, LFNCBR89-NCCS may be just 3 or up to 7 times more effort than PW92. Step functions in the functional form give rise to numerical instabilities that lead to larger deviations for SCAN and NCMSCAN. Yet, regarding the total time to solution it is actually more expensive to do SCAN compared to NCMSCAN and LFNCBR89-NCCS. 

In summary, one can say that SCDFT takes roughly 5--7 times the computational effort of LSDA. Therefore, the computational cost should not be an obstacle for most applications.

\section{Results}
\label{sec:Results}

\subsection{Structural properties}
\label{subsec:Structural properties}

\begin{table*}[t]
  \centering
  \caption{
    Relative deviations of the computed equilibrium lattice
    parameters from the $T\to 0$ experimental references of
    Table~\ref{tab:exp lattice}, $\Delta x/x_0=(x_\mathrm{BM}-x_0)/x_0$, and the
    bulk modulus $B_\mathrm{BM}$ (GPa) from the Birch--Murnaghan fits.  For
    \MnGe\ the MAGNDATA room-temperature structure is the reference, since
    low-temperature data exist only for Mn-rich Mn$_{3.2}$Ge; because the
    fits scale the cell isotropically at fixed $c/a$,
    $\Delta c/c_0=\Delta a/a_0$ for \MnGe.  Also listed are the average
    on-site magnetic moment $m$ per magnetic ion (spin only;
    spin+orbital for \NiS) and, for the insulators, the fundamental gap
    $E_g$, both evaluated at the sampled EOS volume closest to the BM
    minimum; negative gaps (band overlap) are shown as 0, together with
    their relative deviations $\Delta m/m_0$ and $\Delta E_g/E_{g,0}$ where
    experimental values exist.  The last column
    repeats the experimental reference values of Table~\ref{tab:exp lattice}.}
  \label{tab:lattice}
  \begin{ruledtabular}
  \begin{tabular}{llrrrrrrr}
& & PZ & PW92 & PBE & SCAN & NCMSCAN & LFNCBR89-NCCS & expt \\
    \colrule
           & $a_\mathrm{BM}$ (\AA)  & 3.572 & 3.576 & 3.716 & 3.738 & 3.591 & 3.729 & 3.782 \\
    \MnIr  & $\Delta a/a_0$ (\%)    & $-5.55$ & $-5.44$ & $-1.76$ & $-1.16$ & $-5.06$ & $-1.39$ & --- \\
           & $\Delta V/V_0$ (\%)    & $-15.74$ & $-15.44$ & $-5.18$ & $-3.45$ & $-14.43$ & $-4.12$ & --- \\
           & $B_\mathrm{BM}$ (GPa)  & 255 & 237 & 158 & 153 & 236 & 171 & --- \\
           & $m$ ($\muB$) & 1.46 & 1.65 & 2.69 & 3.44 & 1.66 & 3.54 & --- \\
    \colrule
           & $a_\mathrm{BM}$ (\AA)  & 5.017 & 5.014 & 5.246 & 5.356 & 5.072 & 5.321 & 5.302 \\
           & $c_\mathrm{BM}$ (\AA)  & 4.058 & 4.056 & 4.243 & 4.333 & 4.103 & 4.305 & 4.289 \\
    \MnGe  & $\Delta a/a_0$ (\%)    & $-5.38$ & $-5.43$ & $-1.06$ & $+1.02$ & $-4.33$ & $+0.36$ & --- \\
           & $\Delta c/c_0$ (\%)    & $-5.38$ & $-5.43$ & $-1.06$ & $+1.02$ & $-4.33$ & $+0.36$ & --- \\
           & $\Delta V/V_0$ (\%)    & $-15.29$ & $-15.43$ & $-3.15$ & $+3.09$ & $-12.45$ & $+1.09$ & --- \\
           & $B_\mathrm{BM}$ (GPa)  & 177 & 160 & 112 & 89 & 195 & 114 & --- \\
           & $m$ ($\muB$) & 1.44 & 1.43 & 2.43 & 3.33 & 1.50 & 3.53 & 2.65(2) \\
           & $\Delta m/m_0$ (\%)    & $-45.63$ & $-46.19$ & $-8.43$ & $+25.79$ & $-43.34$ & $+33.33$ & --- \\
    \colrule
           & $a_\mathrm{BM}$ (\AA)  & 5.490 & 5.490 & 5.636 & 5.683 & 5.607 & 5.437 & 5.689 \\
    \NiS   & $\Delta a/a_0$ (\%)    & $-3.50$ & $-3.50$ & $-0.94$ & $-0.11$ & $-1.44$ & $-4.43$ & --- \\
           & $\Delta V/V_0$ (\%)    & $-10.13$ & $-10.14$ & $-2.79$ & $-0.33$ & $-4.25$ & $-12.70$ & --- \\
           & $B_\mathrm{BM}$ (GPa)  & 142 & 143 & 103 & 100 & 121 & 152 & --- \\
           & $m$ ($\muB$) & 0.00 & 0.00 & 0.70 & 1.44 & 0.95 & 1.11 & 0.98(6) \\
           & $\Delta m/m_0$ (\%)    & $-100.00$ & $-100.00$ & $-28.60$ & $+46.52$ & $-3.32$ & $+13.47$ & --- \\
           & $E_g$ (eV)             & 0 & 0 & 0 & 0.49 & 0 & 0 & 0.3 \\
           & $\Delta E_g/E_{g,0}$ (\%) & $-100.00$ & $-100.00$ & $-100.00$ & $+61.89$ & $-100.00$ & $-100.00$ & --- \\
    \colrule
           & $a_\mathrm{BM}$ (\AA)  & 6.039 & 6.025 & 6.170 & 6.131 & 6.118 & 6.014 & 6.115 \\
           & $c_\mathrm{BM}$ (\AA)  & 11.188 & 11.161 & 11.430 & 11.359 & 11.334 & 11.142 & 11.367 \\
    \YMO   & $\Delta a/a_0$ (\%)    & $-1.25$ & $-1.48$ & $+0.90$ & $+0.27$ & $+0.05$ & $-1.65$ & --- \\
           & $\Delta c/c_0$ (\%)    & $-1.58$ & $-1.81$ & $+0.56$ & $-0.07$ & $-0.29$ & $-1.98$ & --- \\
           & $\Delta V/V_0$ (\%)    & $-4.02$ & $-4.69$ & $+2.37$ & $+0.47$ & $-0.19$ & $-5.20$ & --- \\
           & $B_\mathrm{BM}$ (GPa)  & 191 & 174 & 148 & 183 & 179 & 201 & 112(8) \\
           & $m$ ($\muB$) & 3.32 & 3.34 & 3.47 & 3.56 & 3.41 & 3.60 & 2.90(2) \\
           & $\Delta m/m_0$ (\%)    & $+14.64$ & $+15.24$ & $+19.61$ & $+22.71$ & $+17.54$ & $+24.07$ & --- \\
           & $E_g$ (eV)             & 0.70 & 0.53 & 0.79 & 1.48 & 0.97 & 1.03 & 1.5 \\
           & $\Delta E_g/E_{g,0}$ (\%) & $-53.55$ & $-64.38$ & $-47.54$ & $-1.59$ & $-35.47$ & $-31.49$ & --- \\
  \end{tabular}
  \end{ruledtabular}
\end{table*}

The BM EOS for each material and XC functional is shown in Figs.~\ref{fig:eos}(e)--(h). Table~\ref{tab:lattice} presents the corresponding relative deviation of the BM minimum with respect to the experimental values $a_0$, $c_0$, and $V_0$ listed in Table~\ref{tab:exp lattice}. With relative deviations of $\le 1.2\%$ for the equilibrium lattice constant, SCAN can predict the experimental value most accurately and shows no systematic over- or underestimation within the four systems considered here. The two LSDA functionals, on the other hand, clearly show the known underestimation of the equilibrium lattice constant. PBE performs similarly to SCAN, but with a larger variance, and it fails to capture the localized nature of the Ni-$d$ bands, as discussed in more detail in Sec.~\ref{subsec:band-structure case study on nis}.

The SCDFT functionals, LFNCBR89-NCCS and NCMSCAN, show very different tendencies: 
The structural properties of the two metallic compounds featuring itinerant magnetism, \MnIr\ and \MnGe\ in Figs.~\ref{fig:eos}(e) and (f), are relatively well described by LFNCBR89-NCCS, while NCMSCAN significantly underestimates the equilibrium volume. For the charge-transfer insulator \YMO,\cite{varignon2022dependence} LFNCBR89-NCCS underestimates the equilibrium lattice constant to the same degree as the LSDA functionals, while NCMSCAN reproduces the experimental lattice parameters. The difference must originate from the formulation of the XC energy of LFNCBR89-NCCS and NCMSCAN.
All tested functionals overestimate the bulk modulus for \YMO\ even when the volume reproduces the experimental value. This is partly because the experimental measurement was performed at 298\,K, and partly because our EOS scans keep the internal coordinates and $c/a$ frozen, which stiffens the lattice response. Overall, the predicted bulk moduli show a large spread across functionals and materials, which sets the functionals apart further.

From Fig.~\ref{fig:eos}(g), it is clear that the different XC functionals are struggling to describe the structural properties of the Mott insulator \NiS. We therefore discuss the band structure and structural properties in more detail in Sec.~\ref{subsec:band-structure case study on nis}.  

\subsection{On-site magnetic moments and band gaps}
\label{subsec:moments and gaps}

Table~\ref{tab:lattice} also lists the on-site magnetic moments $m$ and, for the insulators, the fundamental band gaps $E_g$, both evaluated at each functional's equilibrium volume. We recall that the quality of the experimental reference differs greatly between the four materials: the on-site moment is precisely known for \MnGe\ [$2.65(2)\,\muB$ from spherical neutron polarimetry\cite{soh2020ground}], \YMO\ [$2.90(2)\,\muB$\cite{munoz2000magnetic}], and \NiS\ [$0.98(6)\,\muB$\cite{yano2016magnetic}], whereas for \MnIr\ no reliable measurement exists for reasons discussed in Appendix~\ref{app:Structural parameters}.

For the two itinerant magnets, the computed moments split the functionals into the same two groups as the EOS analysis: PZ, PW92, and NCMSCAN yield $1.43$--$1.5\,\muB$ for \MnGe\ ($1.46$--$1.66\,\muB$ for \MnIr), while SCAN and LFNCBR89-NCCS yield $3.33$--$3.53\,\muB$ ($3.44$--$3.54\,\muB$), with PBE in between at $2.43\,\muB$ ($2.69\,\muB$). \MnGe\ is the one material where experiment can arbitrate between the groups, and the verdict is sobering: the measured $2.65(2)\,\muB$ falls between them. PBE comes closest ($-8\%$), SCAN and LFNCBR89-NCCS overestimate by $26$--$33\%$, and the small-moment group underestimates by roughly $45\%$, although the Mn-rich off-stoichiometry of real samples discussed in Appendix~\ref{app:Structural parameters} remains a caveat. We further note that the small-moment functionals are exactly those that compress the equilibrium volumes by $12$--$15\%$, cf.~Table~\ref{tab:lattice}; the interplay between moment formation and volume via the magnetovolume effect is taken up again in Sec.~\ref{subsec:MAE}.

The localized magnet \YMO\ presents the opposite picture: all six functionals yield $3.32$--$3.60\,\muB$, i.e.~a spread of merely $0.28\,\muB$ compared to $>2\,\muB$ for the itinerant systems, and all overestimate the measured ordered moment of $2.90(2)\,\muB$ by $14$--$24\%$. This uniform overestimation is, however, expected: the measured ordered moment of \YMO\ is itself strongly reduced from the ionic $4\,\muB$ of Mn$^{3+}$ ($S=2$) by geometric frustration, and inelastic neutron scattering shows that strong spin fluctuations persist deep in the ordered phase.\cite{sato2003unconventional} This suggests that the experimentally measured magnetic moment is reduced by quantum spin fluctuations, which lie beyond any ground-state, time-independent DFT description.

\NiS\ sits between these two limits: the ionic $2\,\muB$ of Ni$^{2+}$ ($S=1$) is roughly halved by the strong hybridization with the sulfur dimers, making the moment a sensitive probe of how each functional balances localization against covalency. Plain NCMSCAN ($0.95\,\muB$) lands within the experimental uncertainty and LFNCBR89-NCCS ($1.11\,\muB$) within $13\%$ of the measured $0.98(6)\,\muB$, PBE underestimates ($0.70\,\muB$), SCAN overestimates ($1.44\,\muB$), and the two LSDAs collapse to a nonmagnetic solution altogether. The electronic structure behind these numbers is dissected in Sec.~\ref{subsec:band-structure case study on nis}.

Regarding the band gaps, experimental references exist for the two insulators. For \NiS\ ($E_g\approx0.3$\,eV\cite{sato1984reflectivity}), all functionals except SCAN ($0.49$\,eV) yield a metallic solution at their equilibrium volume, cf.~Sec.~\ref{subsec:band-structure case study on nis} for the role of an on-site $U$. For \YMO\ ($E_g\approx1.5$\,eV\cite{kalashnikova2003electronic}), only SCAN ($1.48$\,eV) reaches the experimental charge-transfer gap, the SCDFT functionals recover about two-thirds of it ($0.97$ and $1.03$\,eV), and the LSDAs and PBE fall below $0.8$\,eV.

In summary, the spread among the functionals is evident. The band gaps single out SCAN, although this comes at the price of, at minimum, an overestimated magnetic moment. As we show in Sec.~\ref{subsec:Magnetic energy landscape}, the price is even higher: SCAN also misidentifies the magnetic ground state. Because it would be too much to discuss the plethora of phenomena relevant for the four benchmark materials, we now zoom in on two aspects where the functionals disagree qualitatively, namely the electronic structure of the Mott insulator \NiS\ and the magnetic energy landscape and anisotropy of \MnIr, and leave the remaining phenomena to future work.

\subsection{Case study on \NiS}
\label{subsec:band-structure case study on nis}

\begin{figure*}[t]
  \centering
  \includegraphics[width=0.95\textwidth]{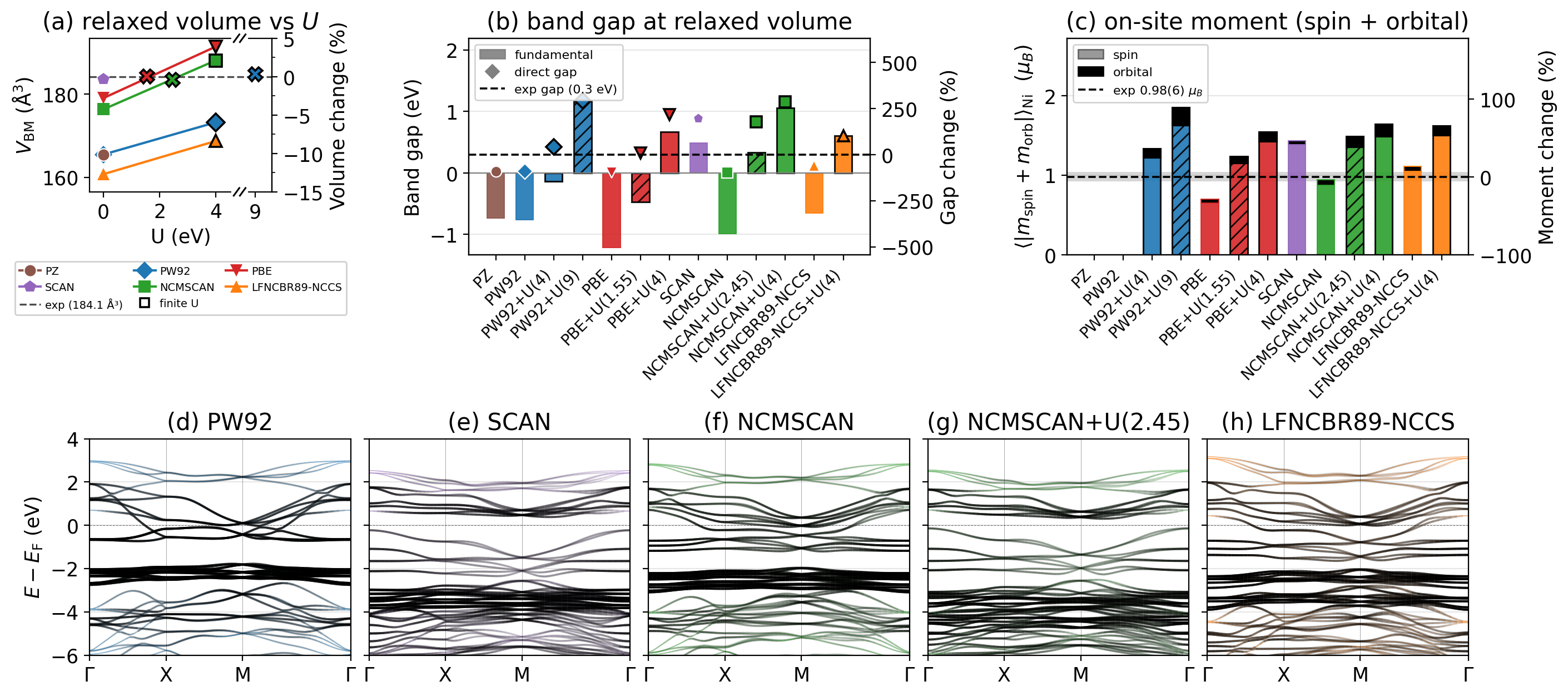}
  \caption{Structural, electronic, and magnetic properties of \NiS\ depending on the exchange-correlation (XC) functional and on-site Coulomb interaction $U$.
  (a)~Equilibrium volume $V_\mathrm{BM}$ from the Birch--Murnaghan fits as
  a function of the on-site $U$ applied to the Ni-$3d$ shell: all
  functionals at $U=0$; PW92, PBE, NCMSCAN, and LFNCBR89-NCCS at $U=4$\,eV
  (black-outlined markers); and NCMSCAN at $U=2.45$\,eV (cross), the value
  chosen such that $V_\mathrm{BM}$ matches the experimental
  low-temperature volume of $184.1$\,\AA$^3$ (2\,K,
  Ref.~\onlinecite{yano2016magnetic}; dashed line). The right axis gives
  the volume change relative to this reference.
  (b)~Fundamental (bars) and smallest direct (markers) band gap at each XC
  functional's $V_\mathrm{BM}$. Negative values denote the band overlap
  of a metallic solution. The dashed line marks the experimental band
  gap of $\approx 0.3$\,eV.\cite{sato1984reflectivity}
  (c)~Ni on-site moment including spin and orbital contributions at $V_\mathrm{BM}$: the
  colored bar is the spin part, the black segment on top the orbital
  contribution.  The dashed line is the experimental ordered moment of
  $0.98(6)\,\muB$ (2\,K, MAGNDATA entry~0.150).\cite{yano2016magnetic}
  PZ and PW92 converge to a nonmagnetic state at $V_\mathrm{BM}$.
  (d)-(h)~Band structures along $\Gamma$-X-M-$\Gamma$ at the
  respective $V_\mathrm{BM}$ for PW92, SCAN, NCMSCAN,
  NCMSCAN+$U$(2.45\,eV), and LFNCBR89-NCCS, relative to the Fermi level. The
  opacity of the black overlay is proportional to the Ni-$d$ orbital character, while the colored lines trace all bands.}
  \label{fig:NiS}
\end{figure*}

To account for the strong electronic correlation in the Mott insulating phase of \NiS, we additionally include an on-site Coulomb interaction $U$ within the standard DFT+$U$\cite{dudarev1998electron} method for PW92, PBE, NCMSCAN, and LFNCBR89-NCCS. Figure~\ref{fig:NiS}(a) shows the equilibrium volume $V_{\mathrm{BM}}$ as a function of $U$. All functionals show the same qualitative trend that increasing $U$ compresses the Ni-$d$ orbitals. The Ni-$sp$ orbitals hybridize with the $d$-states and expand accordingly. The expansion of Ni-$sp$ orbitals leads to an increase in the lattice parameter as known from NiO.\cite{dudarev1998electron}

A simple phenomenological approach to determine the suitable $U$ is to require that the equilibrium volume matches experiment. Figure~\ref{fig:NiS}(a) shows that this leads to reasonable $U$ values for PBE and NCMSCAN, while PW92 and LFNCBR89-NCCS would require exceedingly large values for $U$. We should recall here that SCAN already reproduces the experimental volume without $U$. Looking at the fundamental band gap presented in Fig.~\ref{fig:NiS}(b), we see that all $U=0$\,eV calculations, except for SCAN, are metallic. PW92+$U$ with $U=4$\,eV even remains metallic although a direct gap opens. This semi-metallic behavior is already present for LFNCBR89-NCCS at $U=0$\,eV. Moreover, despite the similar volume trend of PW92 and LFNCBR89-NCCS, the fundamental band gap already exceeds the experimental band gap of $\approx0.3$\,eV\cite{sato1984reflectivity} for LFNCBR89-NCCS+$U$ at $U=4$\,eV. Hence increasing $U$ for LFNCBR89-NCCS to reconcile volume, band gap, and magnitude of the moments proves impossible.

For NCMSCAN, we also see the band gap opening by applying a finite value for $U$. A linear interpolation of the relaxed volume at $U=0$\,eV and $U=4$\,eV shows that $U=2.45$\,eV should yield the experimental volume. We perform a full EOS calculation and confirm that the relaxed volume corresponds to the experimental volume within 0.39\%. Interestingly, the additional NCMSCAN+$U$ calculation at $U=2.45$\,eV reproduces the experimental band gap, cf.~Fig.~\ref{fig:NiS}(b). Unfortunately, this predictive power does not extend to the on-site magnetic moment: Considering the on-site magnetic moment including both the spin and orbital contribution presented in Fig.~\ref{fig:NiS}(c) shows that NCMSCAN and LFNCBR89-NCCS closely agree with the experimental value at $U=0$\,eV. However, the two calculations SCAN and NCMSCAN+$U$(2.45), which are closest to experiment regarding volume and band gap, both overestimate the on-site magnetic moment to a similar degree (SCAN $+46\%$, NCMSCAN+$U$(2.45) $+52\%$).
The same procedure applied to PBE ($U=1.55$\,eV) and PW92 ($U=9$\,eV) also fails to reconcile volume, band gap, and magnitude of the moments. PBE+$U$ at $U=1.55$\,eV stays metallic and PW92+$U$ at $U=9$\,eV grossly overestimates the band gap and magnitude of the moments.

Finally, Figs.~\ref{fig:NiS}(d)--(h) present selected band structures. The Ni-$d$ orbital character is highlighted in black on top of the remaining bands that are plotted in the color of the base functional. We see that the weight of the Ni-$d$ bands is most concentrated near the Fermi level for PW92 and this is in fact similar for PZ and PBE. All three of those yield metallic behavior. Interestingly, NCMSCAN and LFNCBR89-NCCS, presented in Figs.~\ref{fig:NiS}(f) and (h), move the Ni-$d$ states away from the Fermi level and tend towards semi-metallic behavior, i.e.~a direct gap is forming; however, the bands still cross the Fermi energy. Finally, SCAN and NCMSCAN+$U$(2.45), presented in Figs.~\ref{fig:NiS}(e) and (g), yield a very similar dispersion.

This demonstrates that DFT+$U$ can be applied within SCDFT without additional effort, with an effect similar to that for known DFT functionals. It shows that SCAN and NCMSCAN+$U$(2.45) are able to reproduce the electronic properties observed experimentally. Regarding the on-site moments, however, SCAN and NCMSCAN+$U$(2.45) both suffer from overestimation. Here, plain LFNCBR89-NCCS and NCMSCAN give better predictions for the on-site moments. It is also noteworthy that LSDA could not stabilize a magnetic solution without on-site Coulomb interaction $U$ and instead converged to a nonmagnetic state. All calculations thus far are obtained using the experimental magnetic configuration as the initial guess for the on-site moments. In the next section we want to take a closer look at the relative stability of the magnetic configurations for \NiS, as well as for \MnIr.

\subsection{Magnetic energy landscape}
\label{subsec:Magnetic energy landscape}

\begin{figure}[t]
  \centering
  \includegraphics[width=\columnwidth]{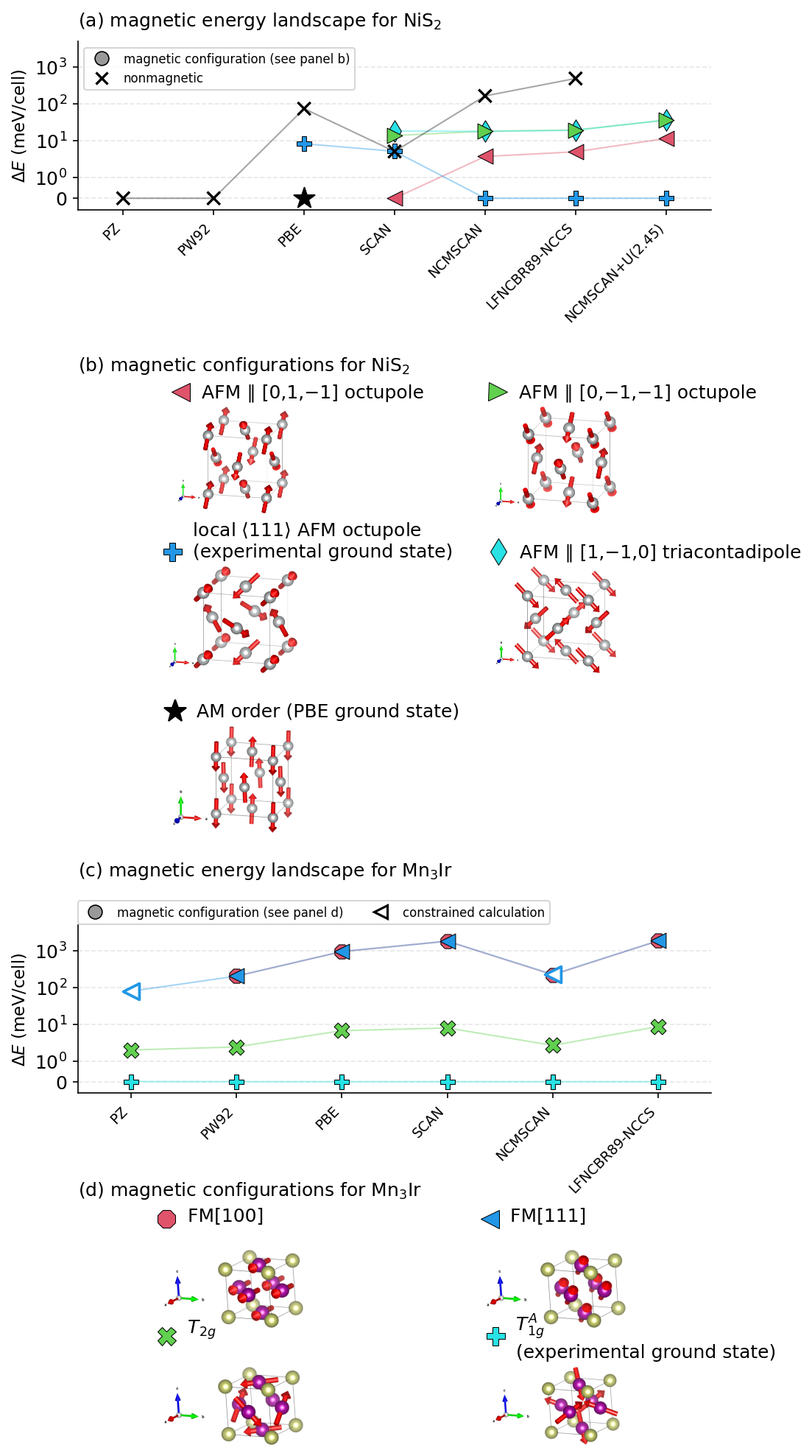}
  \caption{Relaxed-volume magnetic energy landscapes and candidate
    configurations for \NiS\ [(a),(b)] and \MnIr\ [(c),(d)]. In the
    landscape panels (a),(c), $\Delta E$ per cell is shown relative to
    each functional's ground state on a logarithmic scale. The colored
    markers are the magnetic configurations rendered in panels (b),(d),
    black $\times$ marks the nonmagnetic solution, and open $\triangleleft$
    denotes constrained calculations (also shown in Fig.~\ref{fig:MAE}(b) at $\theta=90^\circ$).  The black star
    in (a) is the collinear altermagnetic (AM) order that the PBE ground
    state converges to. Panels (b),(d) show the corresponding magnetic configurations, with the antiferromagnetic (AFM) experimental ground states highlighted.}
  \label{fig:landscape}
\end{figure}

Both in nature and in ab initio calculations, different magnetic orders are known to coexist as local minima yielding metastable configurations. Hence, the initialized magnetic moments strongly bias toward a particular magnetic solution at the end of the SCF calculation. As discussed in Sec.~\ref{subsec:mag order}, we construct different magnetic configurations as starting points to explore the (SC)DFT total energy landscape. 

Figure~\ref{fig:landscape}(a) presents the total energy difference for \NiS\ of each magnetic configuration given in Fig.~\ref{fig:landscape}(b) with respect to the magnetic ground-state configuration for a given XC functional. We have already seen in the preceding section that the LSDA functionals yield a nonmagnetic solution for \NiS\ starting from the experimental magnetic configuration. And in fact, starting from any of the magnetic configurations shown in Fig.~\ref{fig:landscape}(b), the solution remains nonmagnetic, cf.~Fig.~\ref{fig:landscape}(a) black cross.

Most remarkably, LFNCBR89-NCCS and NCMSCAN predict the experimental magnetic order to be most stable among the investigated configurations, while PBE and SCAN find other magnetic ground states. In particular, for PBE, all initial configurations---except for the local $\langle 111\rangle$ AFM octupole---converge to the altermagnetic (AM) order, marked by a black star in Fig.~\ref{fig:landscape}(a). Considering the point group of the crystal structure, the AM order is an unusual combination of three irreducible representations.\cite{huebsch2021benchmark} However, as Park~\emph{et al.}\cite{park2026impact} have argued, the AM order is actually the collinear version of the local $\langle 111\rangle$ AFM octupole. The noncollinearity emerges due to the SOC that aligns the moments along the local $\langle 111\rangle$ axis. Since NCMSCAN+$U$(2.45), like SCAN, reproduces the experimental volume and roughly the band gap, while overestimating the magnitude of the moments (cf.~Sec.~\ref{subsec:band-structure case study on nis}), we also probe its magnetic energy landscape. We find that it stabilizes the existing magnetic configurations, e.g., moving the first metastable solution from 4\,meV/cell up to 11.6\,meV/cell, while destabilizing the nonmagnetic solution. In particular, the nonmagnetic initialization finds a symmetry-equivalent domain of the local $\langle 111\rangle$ AFM octupole shown in Fig.~\ref{fig:landscape}(b). This is in line with an earlier observation in the context of PBE+$U$ in Ref.~\onlinecite{huebsch2021benchmark}, where $+U$ calculations do not tend to change the most stable magnetic configuration.

One may also wonder about the energy difference between the AFM $\parallel$ $[0,1,-1]$ octupole and the AFM $\parallel$ $[0,-1,-1]$ octupole, red left-pointing and green right-pointing triangles in Fig.~\ref{fig:landscape}(b), respectively. These are identical noncollinear $\langle 011\rangle$ octupole patterns on the Ni fcc sublattice with sites permuted Ni1$\leftrightarrow$Ni3, Ni2$\leftrightarrow$Ni4. This is exactly a $t=(0,\frac{1}{2},\frac{1}{2})$ shift of the magnetic moments relative to the S octahedra. The tilt of the S octahedra, however, obeys a nonsymmorphic glide operation that involves both the translation $t$ and a mirror reflection on a $\{001\}$ plane. In other words, only the tilt of the S octahedra makes these two AFM orders distinct, and their energy differences represent a measure for the spin-pattern-vs-S-sublattice coupling referred to as AM anisotropy in the following. While the AM anisotropy for SCAN is $14.0$\,meV/cell, it is $14.6$\,meV/cell (NCMSCAN) and $14.7$\,meV/cell (LFNCBR89-NCCS) for the SCDFT functionals. Since PBE does not stabilize these two magnetic configurations we cannot report a value for the AM anisotropy. 
Note that, since SCAN's reference state (its own ground state) is not the experimental order, the distances on the logarithmic axis in Fig.~\ref{fig:landscape}(a) are not directly comparable to those of the other functionals.

In summary, the magnetic landscape for \NiS\ substantially changes depending on the XC functional. Only LFNCBR89-NCCS and NCMSCAN predict the experimental magnetic ground state among the explored functionals. Compared to that, the magnetic energy landscape for \MnIr\ shown in Figs.~\ref{fig:landscape}(c) and (d) looks less diverse: All XC functionals predict the same magnetic ground state and order of stability. We remark that some of the FM states are exceedingly hard to converge and thus we used constrained-moments calculations as discussed in Sec.~\ref{subsec:mag order} and highlighted by empty markers in Fig.~\ref{fig:landscape}(c). Surprisingly, upon closer investigation, the magnetic states obtained via the different XC functionals for \MnIr\ are quite distinct when mapping onto an effective spin Hamiltonian. This is explored in the subsequent section.

\subsection{Magnetic anisotropy energy of \MnIr}
\label{subsec:MAE}

\begin{table*}[t]
  \centering
  \caption{
    Magnetic anisotropy and exchange parameters of \MnIr\ per cell
    extracted from the fits in Fig.~\ref{fig:MAE}: average on-site Mn moment
    $|m|$ in the $T_{1g}^A$ ground state, in-plane anisotropy
    $K_\mathrm{eff}$, exchange stiffness $K_\mathrm{FM}$, normalized
    exchange constant $J=K_\mathrm{FM}/(6|m|^2)$, and normalized
    anisotropy $K=K_\mathrm{eff}/(2|m|^2)$.
    In the underlying Heisenberg mapping of
    Ref.~\onlinecite{mccoombs2023impact},
    $H=\sum_{\langle ij\rangle}J\,\mathbf{S}_i\cdot\mathbf{S}_j
      - K \sum_i \big(\hat n_i\!\cdot\!\mathbf{S}_i\big)^2$,
    with $\hat n_i$ the local $\langle100\rangle$ anisotropy axis of
    sublattice $i$, the spin vectors carry the moment magnitude in
    $\muB$, hence $J$ and $K$ are quoted in meV/$\muB^2$.
    For conversion into anisotropy-energy densities: at the experimental
    cell volume $V_0=54.1$\,\AA$^3$, 1\,meV per cell corresponds to
    $18.5\,\mu$eV/\AA$^3=2.96$\,MJ/m$^3=2.96\times10^{7}$\,erg/cm$^3$;
    note, however, that each functional's value must be converted using
    its own equilibrium volume listed in Table~\ref{tab:lattice}.}
  \label{tab:MAE}
  \begin{ruledtabular}
  \begin{tabular}{lcccccc}
     & PZ & PW92 & PBE & SCAN & NCMSCAN & LFNCBR89-NCCS \\
    \colrule
    $|m|$ ($\muB$) & 1.46 & 1.65 & 2.69 & 3.44 & 1.66 & 3.54 \\
    $K_\mathrm{FM}$ (meV/cell) & 422 & 572 & 1464 & 1857 & 580 & 1845 \\
    $K_\mathrm{eff}$ (meV/cell) & 2.0 & 2.4 & 6.7 & 8.0 & 2.7 & 8.7 \\
    $J$ (meV/$\muB^2$) & 33.1 & 35.2 & 33.7 & 26.1 & 35.3 & 24.5 \\
    $K$ (meV/$\muB^2$) & 0.47 & 0.45 & 0.46 & 0.34 & 0.49 & 0.35 \\
  \end{tabular}
  \end{ruledtabular}
\end{table*}

At first sight, the magnetic energy landscape of \MnIr\ in
Fig.~\ref{fig:landscape}(c) suggests that the choice of the XC functional is uncritical: all six functionals find the $T_{1g}^A$ ground state and agree on the energetic ordering of the metastable configurations. A more sensitive probe is the magnetic anisotropy energy (MAE), which provides a first hint at the spin dynamics the system exhibits. Rotating all Mn moments by an angle $\varphi$ about the $[111]$ axis transforms $T_{1g}^A$ ($\varphi=0$) into the $T_{2g}$ configuration ($\varphi=90^\circ$) while preserving the relative $120^\circ$ angles between the moments.

An exceptional stiffness of the AFM order motivates the use of \MnIr\ as the pinning layer in exchange-bias devices.\cite{parkin2003magnetically} Korringa--Kohn--Rostoker (KKR) LSDA calculations\cite{szunyogh2009giant} for \MnIr\ predict a large effective anisotropy of $K_\mathrm{eff}=10.42$\,meV/cell in line with earlier LSDA estimates.\cite{sakuma2003first} 
For the sputtered, chemically disordered $\gamma$-phase films employed there, grain-volume analyses of blocking-temperature distributions yield anisotropy densities of $3.4\,\mu$eV/\AA$^3$ at room temperature\cite{vallejo2007measurement} rising to $17\,\mu$eV/\AA$^3$ for optimally textured films.\cite{aley2008texture} This is one to two orders of magnitude below the KKR-LSDA prediction of $\approx190\,\mu$eV/\AA$^3$ for the ordered L1$_2$ phase.\cite{szunyogh2009giant} For the ordered phase itself no measurement exists. 

McCoombs~\emph{et al.}\cite{mccoombs2023impact} presented a detailed symmetry analysis where they showed that the general bilinear spin Hamiltonian of this lattice additionally admits a staggered Dzyaloshinskii--Moriya interaction and an off-diagonal ($\Gamma$-type) anisotropic exchange. These contributions, however, cancel identically in all $\mathbf q=0$ spin configurations investigated here. Hence, they are invisible to the presented total-energy calculations and appear only in the spin-wave dispersion, which is beyond the scope of this work.

\begin{figure}[t]
  \centering
  \includegraphics[width=\columnwidth]{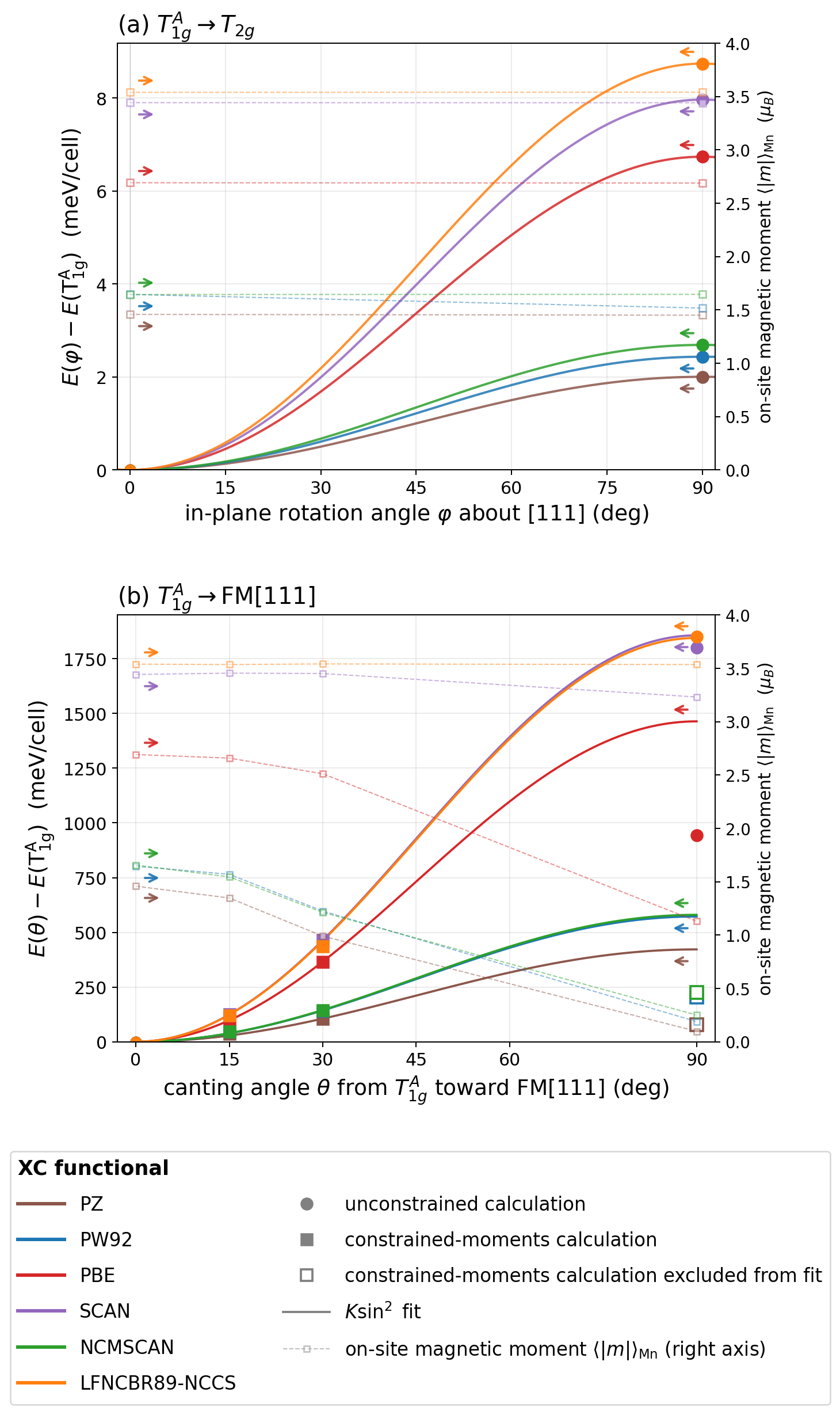}
  \caption{Magnetic anisotropy of \MnIr\ along two paths out of the
    $T_{1g}^A$ ground state resolved by XC functional (colors).
    (a)~In-plane rotation by an angle $\varphi$ about the $[111]$ axis,
    connecting $T_{1g}^A$ ($\varphi=0$) to $T_{2g}$ ($\varphi=90^\circ$):
    the energy difference $E(\varphi)-E(T_{1g}^A)$ per cell (left axis) follows a
    $K_\mathrm{eff}\sin^2\varphi$ law (solid lines). (b)~Canting by an angle $\theta$
    from $T_{1g}^A$ toward FM$[111]$, evaluated with constrained-moment
    calculations (squares) and fitted by $K_\mathrm{FM}\sin^2\theta$ (using filled squares; omitting open markers). In both panels, dashed curves with open squares give the on-site Mn moment
    $\langle|m|\rangle_{\rm Mn}$ read on the right axis.}
  \label{fig:MAE}
\end{figure}

Figure~\ref{fig:MAE} shows the energy along two paths out of the $T_{1g}^A$ ground state, evaluated at each functional's equilibrium volume $V_\mathrm{BM}$. For the in-plane rotation, presented in Fig.~\ref{fig:MAE}(a), $E(\varphi)$ is fitted by the single-parameter form $K_\mathrm{eff}\sin^2\varphi$. 
All functionals yield $K_\mathrm{eff}>0$, but its magnitude splits them into two groups, cf.~Table~\ref{tab:MAE}: PBE, SCAN, and LFNCBR89-NCCS give values in the range of $6.7$--$8.7$\,meV/cell, whereas PZ, PW92, and NCMSCAN yield roughly three times smaller ones. Expressed as an energy density at the respective equilibrium volumes, $K_\mathrm{eff}$ corresponds to $44$--$169\,\mu$eV/\AA$^3$. In other words, all functionals place the ordered phase a factor of 3--10 above the measured anisotropy of the disordered $\gamma$-films quoted above.

This grouping correlates exactly with the on-site Mn moments shown on the right axis of Fig.~\ref{fig:MAE}(a). For PBE, SCAN, and LFNCBR89-NCCS, we find $2.7$--$3.5\,\muB$ versus $1.5$--$1.7\,\muB$ for PZ, PW92, and NCMSCAN. It is important to stress here that the experimental value of the Mn moments is, to the best of our knowledge, not reliably known and thus one cannot say which group of functionals is closer to the ground truth.
The reason that the moment of such a widely applied material has remained unmeasured has a practical origin:
Ir has one of the largest thermal-neutron absorption cross sections of all stable elements,\cite{sears1992neutron} which renders neutron experiments on bulk \MnIr\ exceedingly difficult. The single-crystal diffraction experiment of Tomeno~\emph{et al.},\cite{tomeno1999magnetic} on which the MAGNDATA entry is based, unambiguously established the $T_{1g}^A$ order, but the Ir absorption prevented a reliable extraction of the moment magnitude from the integrated intensities. Thin-film samples, as employed in exchange-bias devices, do not offer a way of determining the size of the moment either, since their small uncompensated moments are dominated by extrinsic strain, interface, and grain effects.\cite{taylor2019magnetic}

Although in this light the magnitude of the Mn moments might not be of primary interest, we suspect they have substantial implications as a secondary effect: The equilibrium volumes are also strongly compressed by $\Delta V/V_0\approx-15$\,\% with respect to experiment for PZ, PW92, and NCMSCAN. Hence, invoking the magnetovolume effect (piezomagnetic response), the smaller Mn moments are perhaps the underlying cause of the underestimated lattice constant. We also note that normalizing $K_\mathrm{eff}$ to yield the normalized anisotropy $K=K_\mathrm{eff}/(2|m|^2)$ collapses the observed spread. All values for $K$ are within the range of $0.34$--$0.49$\,meV/$\muB^2$, cf.~Table~\ref{tab:MAE}.

For the second path, shown in Fig.~\ref{fig:MAE}(b), the moments are canted out of the $(111)$ plane by an angle $\theta$ toward the collinear FM$[111]$ configuration, which is the FM partner $T_{1g}^F$ of the ground-state irreducible representation. As the FM$[111]$ is a local maximum in the magnetic energy landscape, we had to employ constrained-moment calculations and perform a $\lambda\to\infty$ extrapolation as described in Sec.~\ref{subsec:mag order}. Depending on the functional, we see that the Mn moments drastically decrease for the FM$[111]$ configuration. This is directly correlated with the size of the moments in the $T_{1g}^A$ ground state: SCAN and LFNCBR89-NCCS retain nearly the full moment, PBE loses more than half of it, while the LSDA functionals and NCMSCAN yield virtually nonmagnetic solutions.
For LFNCBR89-NCCS, the $\sin^2\theta$ fit through the constrained points at $\theta\le30^\circ$
reproduces the independent, unconstrained FM$[111]$ energy to better than $1$\,\% ($3$\,\% for SCAN). The rigidity of the moment magnitude for LFNCBR89-NCCS and SCAN means the moments behave as Heisenberg-like effective spins, which is favorable for constructing an effective spin model. However, there is no experimental evidence that the moments are Heisenberg-like, for the same reason that the Mn moments are difficult to measure in the first place, namely the neutron absorption of Ir.

To extract the exchange stiffness $K_\mathrm{FM}$ for the ground-state $T_{1g}^A$ configuration, we perform additional constrained-moment calculations at $\theta=15^\circ$ and $30^\circ$, and the energies are fitted by $K_\mathrm{FM}\sin^2\theta$ taking into account only the data points shown with filled markers in Fig.~\ref{fig:MAE}(b). Rotating the moments toward FM$[111]$ gives up the frustrated AFM exchange, and $K_\mathrm{FM}$ is therefore dominated by the isotropic exchange coupling.  The fitted values, listed in Table~\ref{tab:MAE}, split into the same two groups, separated by roughly a factor of four. Again, normalizing by the square of the on-site moment, $J=K_\mathrm{FM}/(6|m|^2)$, collapses this spread: the exchange stiffness per moment is nearly independent of the functional, and the spread in the raw energies is chiefly a moment-magnitude effect.

\section{Conclusion}
\label{sec:Conclusion}

In summary, we have compared two SCDFT functionals, NCMSCAN and LFNCBR89-NCCS, to the locally collinear versions of (semi-)local functionals (PZ, PW92, PBE, and SCAN) using VASP on a small but diverse set of noncollinear $d$-electron magnets: the itinerant triangular antiferromagnets \MnIr\ and \MnGe, the noncollinear narrow-gap Mott insulator \NiS, and the multiferroic insulator \YMO. 

Among the magnetic configurations investigated, only the two SCDFT functionals recover the experimental noncollinear spin texture of \NiS.
PBE instead stabilizes a collinear altermagnetic order, SCAN selects a different magnetic order, and the two LSDAs (PZ, PW92) lead to nonmagnetic solutions.

The SCDFT functionals also reproduce the ordered moment of NiS$_2$ more closely than any other functional tested. However, this qualitative success is not matched by uniform quantitative accuracy: SCAN remains the most reliable for equilibrium volumes and band gaps, while the SCDFT functionals show material-class-specific volume errors of up to 15\% compression. Here it seems the challenge for an XC approximation is to encode both itinerant magnetism and localized magnetism. For \NiS\ we also found that no on-site Coulomb correction $U$ reconciles volume, band gap, and magnetic moment for any functional investigated.

Moreover, the on-site magnetic moment emerges as the variable that organizes much of the disagreement between the functionals. 
For the itinerant magnets \MnIr\ and \MnGe, PZ, PW92, and NCMSCAN form a small-moment group ($1.4$--$1.7\,\muB$ on Mn) that also compresses the equilibrium volumes, consistent with a magnetovolume effect, whereas SCAN and LFNCBR89-NCCS form a large-moment group ($3.3$--$3.5\,\muB$), with PBE in between and closest to the moment measured for \MnGe. 
For \MnIr, a seemingly functional-independent magnetic energy landscape conceals a three- to fourfold spread in the magnetic anisotropy and exchange stiffness. Normalizing by the square of the on-site moment collapses this spread, showing that an error in the moment propagates roughly quadratically into every spin-model parameter extracted from total energies. It serves as a reminder that agreement on the magnetic ground state does not imply agreement on the effective spin Hamiltonian.

The recent application-focused review in Ref.~\onlinecite{marzari2021electronic} framed magnets as a challenging class of materials requiring individual attention for each system studied. Our findings sharpen that statement for the noncollinear case: the difficulty is not merely a sensitivity to parameter selection, since for \NiS\ the locally collinear functionals do not fail by a small margin but converge to qualitatively different spin textures. This shows that going beyond locally collinear functionals can matter decisively for the description of noncollinear magnetism.

Regarding the computational effort, a single SCDFT SCF step costs about seven times an LSDA step and about $1.5$ times an MGGA step, while in the total time to solution the distinction between MGGA and SCDFT is no longer resolvable, cf.~Table~\ref{tab:cost summary}. The computational cost should therefore not be an obstacle for most applications.

Going forward, from a developer's perspective we would hence benefit tremendously from a well-curated and more extensive list of experimentally established values for properties at $T\to0$, e.g.\ equation-of-state fits, band gaps, on-site magnetic moments, magnetic order, magnon spectra, and phonon dispersions for magnets with a small magnetic unit cell. The commendable effort by Gallego and coworkers in establishing MAGNDATA\cite{gallego2016magndata,gallego2016magndata2} is exactly the kind of groundwork that facilitates larger benchmarks of XC functionals and can, thus, help form a systematic understanding of where noncollinear first-principles calculations need to be improved. On the computational side, the density functional approximations may be further benchmarked against semirelativistic configuration-interaction calculations for noncollinear magnetism of small molecules.

It remains to say that the rich world of noncollinear magnetism should be explored further within the framework provided by SCDFT. Even the systems discussed in this paper are not described exhaustively. For instance, the famous homometric pairs in \YMO\ resolved by second-harmonic generation\cite{fiebig2000determination} differ by a pure SOC-driven in-plane anisotropy. Another interesting direction is the magnon dispersion and anomalous transport in \MnGe\cite{cable1993magnetic,sukhanov2019magnon} and Mn$_3$Sn.\cite{ikhlas2017large,park2018magnetic,higo2018large,nomoto2020cluster,chen2021anomalous,tanaka2025ab,tsai2026picosecond} Recent applications to topological insulators, semimetals, and Rashba materials have yielded encouraging results.\cite{comaskey2022role,bodo2022spin,DesmaraisPRL2024,Desmarais2024:PRM:13802,Boccuni2024} On the theoretical side, further opportunities lie in the investigation of spin dynamics within time-dependent SCDFT.

\section*{ACKNOWLEDGMENTS}

J. K. D. was supported by the Project CH4.0 under the “Ministero dell’Università e della Ricerca” (MUR) program “Dipartimenti di Eccellenza 2023-2027” (CUP: D13C22003520001).

\section*{AUTHOR DECLARATIONS}

\subsection*{Conflict of Interest}

Marie-Therese Huebsch, Martijn Marsman, and Fabien Tran are employees of the VASP Software GmbH.

\subsection*{Author Contributions}

\textbf{Marie-Therese Huebsch}: Conceptualization (lead); Software (supporting); Data curation (lead); Investigation (lead); Formal analysis (lead); Methodology (lead); Writing - original draft (lead); Writing - review \& editing (lead).
\textbf{Martijn Marsman}: Conceptualization (supporting); Software (supporting); Writing - review \& editing (supporting).
\textbf{Jacques K. Desmarais}: Conceptualization (supporting); Software (supporting); Investigation (supporting); Methodology (supporting); Writing - original draft (supporting); Writing - review \& editing (supporting).
\textbf{Stefano Pittalis}: Conceptualization (supporting); Investigation (supporting); Methodology (supporting); Writing - original draft (supporting); Writing - review \& editing (supporting).
\textbf{Fabien Tran}: Conceptualization (equal); Software (lead); Investigation (supporting); Methodology (equal); Writing - original draft (supporting); Writing - review \& editing (equal).

\section*{DATA AVAILABILITY}

The data that support the findings of this study are available from the corresponding author upon reasonable request.

\appendix

\section{Notation}
\label{app:Notation}

In this appendix, we start by showing the transformations of Hermitian $2\times2$ matrices. Then, the notation for the two kinds of three-component vectors is explained, which clarifies the quantities used in Sec.~\ref{sec:Methods} and in Fig.~\ref{fig:scf-loop}.

\subsection{Transformations between $n_{\sigma\sigma'}$ and $(n,\vec m\,)$}
\label{app:transformation}

Being a Hermitian $2\times2$ matrix in spinor space, the density matrix $n_{\sigma\sigma'}$ (defined in Fig.~\ref{fig:scf-loop}) can be expanded
in the basis spanned by the unit matrix $\mathbbm{1}$ and the three Pauli matrices $\sigma^a$ with superscript $a$.
The expansion coefficients are the charge density $n$ and the magnetization $\vec m$,
\begin{equation}
n_{\sigma\sigma'}
=\frac{1}{2}\left[n\,\delta_{\sigma\sigma'}
+\sum_{a\in\{x,y,z\}} m^a\,\sigma^{a}_{\sigma\sigma'}\right],
\label{eq:app expansion}
\end{equation}
where $\delta_{\sigma\sigma'}=\mathbbm{1}_{\sigma\sigma'}$. Using $\sum_{\sigma\sigma'}\sigma^{a}_{\sigma'\sigma}\sigma^{b}_{\sigma\sigma'}=2\delta_{ab}$, the inverse transformation reads
\begin{subequations}
\begin{equation}
n=\sum_{\sigma\in\{\uparrow,\downarrow\}} n_{\sigma\sigma},
\end{equation}
\begin{equation}
m^a=\sum_{\sigma\sigma'\in\{\uparrow,\downarrow\}}\sigma^{a}_{\sigma'\sigma}\,n_{\sigma\sigma'},
\end{equation}
\label{eq:app inverse}
\end{subequations}
or, written out component by component:
\begin{subequations}
\begin{equation}
n=n_{\uparrow\uparrow}+n_{\downarrow\downarrow},
\end{equation}
\begin{equation}
m^x=n_{\uparrow\downarrow}+n_{\downarrow\uparrow},
\end{equation}
\begin{equation}
m^y=\imag\left(n_{\uparrow\downarrow}-n_{\downarrow\uparrow}\right),
\end{equation}
\begin{equation}
m^z=n_{\uparrow\uparrow}-n_{\downarrow\downarrow}.
\end{equation}
\end{subequations}
The same transformations apply to $\tau_{\sigma\sigma'}$ and to each of the three direct-space components of $\mathbf{j}_{\sigma\sigma'}$, also defined in Fig.~\ref{fig:scf-loop}.

\subsection{Direct-space and spin-space vectors}
\label{app:two vectors}

Two kinds of three-component vector spaces appear throughout this work, and we distinguish them typographically:

(i) Bold symbols, like $\mathbf{j}$, and the $\nabla$ symbol denote Cartesian vectors in direct space. Their components carry a subscript $\nu$ and the inner product of two such vectors reads
\begin{equation}
\mathbf{a}\cdot\mathbf{b}=\sum_{\nu\in\{x,y,z\}} a_\nu\, b_\nu .
\end{equation}

(ii) Arrows, as in $\vec m$, denote vectors in spin space, and the inner product of two such vectors reads
\begin{equation}
\vec u\cdot\vec v=\sum_{a\in\{x,y,z\}} u^a\, v^a .
\end{equation}

Objects carrying both types of indices, such as the paramagnetic spin-current density $J^a_\nu$, are simultaneously vectors in direct space and in spin space; for these we use an index-free notation combining both accents, as in $\vec{\mathbf{J}}$.
Given these definitions for the inner products, the contracted quantities appearing in this work read
\begin{subequations}
\begin{equation}
|\vec m\,|=\Bigg(\sum_{a} m^a\,m^a\Bigg)^{1/2},
\label{eq:app mod m}
\end{equation}
\begin{equation}
\vec m\cdot\vec\tau_m=\sum_{a} m^a\,\tau^{a}_{m},
\label{eq:app m tau}
\end{equation}
\begin{equation}
|\mathbf{j}|^2=\sum_{\nu} j_\nu\, j_\nu,
\label{eq:app j2}
\end{equation}
\begin{equation}
|\nabla\vec m\,|^2=\sum_{a}\nabla m^a\cdot\nabla m^a
=\sum_{a}\sum_{\nu}\left(\partial_\nu m^a\right)\left(\partial_\nu m^a\right),
\label{eq:app grad m2}
\end{equation}
\begin{equation}
|\vec{\mathbf{J}}|^2=\sum_{a}\sum_{\nu} J^{a}_{\nu}\, J^{a}_{\nu},
\label{eq:app J2}
\end{equation}
\end{subequations}
where all sums run over $\{x,y,z\}$. Note that in $|\nabla\vec m\,|^2$ and $|\vec{\mathbf{J}}|^2$ the indices $\nu$ and $a$ are contracted independently.

\section{Structural parameters}
\label{app:Structural parameters}

In this appendix, we document the experimental data underlying the $T\to0$ reference lattice parameters $a_0$, $c_0$, and $V_0$ of Table~\ref{tab:exp lattice}.

For \MnIr, the lattice constant was measured between 292\,K and 1032\,K, cf.~Fig.~7 in Ref.~\onlinecite{tomeno1999magnetic}. As the $T\to0$ reference we use $a_0=3.782$~\AA, obtained from a quadratic extrapolation of the 292--700\,K data. This is close to the value $a=3.785$~\AA\ on which other theoretical works\cite{szunyogh2009giant,zhang2016giant} have based their calculations, while the MAGNDATA entry 0.108 lists $a=3.770$~\AA.

For \MnGe, the structural parameters of MAGNDATA entry 0.377\cite{soh2020ground} are based on room-temperature data; the composition of the underlying sample is not reported. Since stoichiometric \MnGe\ is not stable in equilibrium, real samples are Mn-rich, typically Mn$_{3.2}$Ge, which features a negative thermal expansion:\cite{sukhanov2018gradual} a Debye--Grüneisen fit with $\theta_D$ fixed at 350\,K yields $a'_0=5.353$~\AA\ and $c'_0=4.343$~\AA. Yet, because our calculations use the idealized stoichiometry, we adopt the room-temperature MAGNDATA parameters as the reference.

For \NiS, Yano~\emph{et al.}\cite{yano2016magnetic} measured both the magnetic order and the structural parameters at 2\,K and ambient pressure, yielding $a=5.689$~\AA, which we adopt as $a_0$; the corresponding MAGNDATA entry lists $a=5.687$~\AA. These values are consistent with the room-temperature reports of $a=5.685$~\AA\ by Fujii~\emph{et al.}\cite{sato1983nis2,fujii1987structural} and $a=5.688$~\AA\ by Tanaka~\emph{et al.}\cite{tanaka1993growth} Fujii~\emph{et al.}\ further report a linear pressure dependence of the lattice constant at $T=300$\,K, from $a=5.685$~\AA\ in the insulating phase at 1\,bar to $a=5.585$~\AA\ in the metallic phase at 49\,kbar.

For \YMO, the lattice parameters of MAGNDATA entry 0.6, listed in Table~\ref{tab:exp lattice}, correspond to the room-temperature values of Ref.~\onlinecite{munoz2000magnetic}. As the $T\to0$ reference we adopt the low-temperature plateau in Fig.~12 of the same reference, $a_0\approx6.115$~\AA\ and $c_0\approx11.367$~\AA.

\section{Magnetic configurations for \MnIr\ and \NiS}
\label{app:Magnetic configurations}

Table~\ref{tab:landscape-configs} presents the magnetic configurations used as initial magnetic configurations for the total-energy-landscape calculations in Sec.~\ref{subsec:Magnetic energy landscape}.

\begin{table*}[t]
\centering
\caption{Crystal structures and magnetic configurations of the energy landscape shown in Fig.~\ref{fig:landscape}(b) for NiS$_2$ and (d) for Mn$_3$Ir.
  The moment components $(m_x, m_y, m_z)$ are given in
  $\muB$, normalized to the experimental on-site moments:
  $0.986\,\muB$ per Ni and, since no measured \MnIr{} moment
  exists, the \MnGe{} value $2.652\,\muB$ per Mn. The
  directions are identical for all functionals, while the magnitudes
  relax self-consistently. The associated irreducible representation (irrep) is labeled in Mulliken notation for the
  parent point group at $\Gamma$ ($m\bar{3}$ for NiS$_2$, $m\bar{3}m$
  for Mn$_3$Ir); the multipole rank follows the CMP analysis.}
\label{tab:landscape-configs}

\vspace{1.0em}
\textbf{NiS$_2$}: pyrite structure, space group $Pa\bar{3}$ (No.~205), Ni on the $4a$ fcc sites, S on $8c$ (with $u = 0.393$).\\[0.4em]
\begin{ruledtabular}
\begin{tabular}{l l l *{12}{l}}
 & & & \multicolumn{12}{c}{moment per Ni site ($\muB$)} \\
\cmidrule(lr){4-15}
 &  &  & \multicolumn{3}{c}{Ni$_1$ $(0,0,0)$} & \multicolumn{3}{c}{Ni$_2$ $(\frac12,\frac12,0)$} & \multicolumn{3}{c}{Ni$_3$ $(0,\frac12,\frac12)$} & \multicolumn{3}{c}{Ni$_4$ $(\frac12,0,\frac12)$} \\
\cmidrule(lr){4-6}\cmidrule(lr){7-9}\cmidrule(lr){10-12}\cmidrule(lr){13-15}
configuration & irrep & multipole & $m_x$ & $m_y$ & $m_z$ & $m_x$ & $m_y$ & $m_z$ & $m_x$ & $m_y$ & $m_z$ & $m_x$ & $m_y$ & $m_z$ \\
\midrule
AFM $\parallel$ $[0,1,-1]$ & $T_g$ & octupole & \,\,0 & \,\,0.697 & -0.697 & \,\,0 & -0.697 & \,\,0.697 & \,\,0 & -0.697 & -0.697 & \,\,0 & \,\,0.697 & \,\,0.697 \\
AFM $\parallel$ $[0,-1,-1]$ & $T_g$ & octupole & \,\,0 & -0.697 & -0.697 & \,\,0 & \,\,0.697 & \,\,0.697 & \,\,0 & \,\,0.697 & -0.697 & \,\,0 & -0.697 & \,\,0.697 \\
local $\langle 111 \rangle$ AFM$^{\ddagger}$ & $A_g$ & octupole & \,\,0.569 & \,\,0.569 & \,\,0.569 & \,\,0.569 & -0.569 & -0.569 & -0.569 & \,\,0.569 & -0.569 & -0.569 & -0.569 & \,\,0.569 \\
AFM $\parallel$ $[1,-1,0]$ & $E_g$ & triacontadipole & \,\,0.697 & -0.697 & \,\,0 & \,\,0.697 & \,\,0.697 & \,\,0 & -0.697 & -0.697 & \,\,0 & -0.697 & \,\,0.697 & \,\,0 \\
\midrule
AM order$^{\dagger}$ & $A_g\!\oplus\!E_g$ & --- & -0.024 & -0.661 & -0.026 & -0.024 & \,\,0.661 & \,\,0.026 & \,\,0.024 & -0.661 & \,\,0.026 & \,\,0.024 & \,\,0.661 & -0.026 \\
\end{tabular}
\end{ruledtabular}
\smallskip
{\footnotesize $^{\dagger}$The PBE ground state: the converged collinear altermagnetic state that the $[1,-1,0]$ initialization relaxes into ($|m| \approx 0.66\,\muB$/Ni; cf.\ the black star in Fig.~\ref{fig:landscape}(a)). $^{\ddagger}$Experimental ground state.}

\vspace{1.0em}

\textbf{Mn$_3$Ir}: Cu$_3$Au (L1$_2$) structure, space group $Pm\bar{3}m$
(No.~221),
Mn on the $3c$ face centers, Ir on $1a$ $(0,0,0)$.\\[0.4em]
\begin{ruledtabular}
\begin{tabular}{l l l *{9}{l}}
 & & & \multicolumn{9}{c}{moment per Mn site ($\muB$)} \\
\cmidrule(lr){4-12}
 &  &  & \multicolumn{3}{c}{Mn$_1$ $(0,\frac12,\frac12)$} & \multicolumn{3}{c}{Mn$_2$ $(\frac12,0,\frac12)$} & \multicolumn{3}{c}{Mn$_3$ $(\frac12,\frac12,0)$} \\
\cmidrule(lr){4-6}\cmidrule(lr){7-9}\cmidrule(lr){10-12}
configuration & irrep & multipole & $m_x$ & $m_y$ & $m_z$ & $m_x$ & $m_y$ & $m_z$ & $m_x$ & $m_y$ & $m_z$ \\
\midrule
FM $\parallel [100]$ & $T_{1g}$ & dipole & \,\,2.652 & \,\,0 & \,\,0 & \,\,2.652 & \,\,0 & \,\,0 & \,\,2.652 & \,\,0 & \,\,0 \\
FM $\parallel [111]$ & $T_{1g}$ & dipole & \,\,1.531 & \,\,1.531 & \,\,1.531 & \,\,1.531 & \,\,1.531 & \,\,1.531 & \,\,1.531 & \,\,1.531 & \,\,1.531 \\
$T_{2g}$ (120$^\circ$ AFM $\perp [111]$) & $T_{2g}$ & octupole & \,\,0 & \,\,1.875 & -1.875 & -1.875 & \,\,0 & \,\,1.875 & \,\,1.875 & -1.875 & \,\,0 \\
$T_{1g}^{A}{}^{\ddagger}$ & $T_{1g}$ & octupole & -2.165 & \,\,1.083 & \,\,1.083 & \,\,1.083 & -2.165 & \,\,1.083 & \,\,1.083 & \,\,1.083 & -2.165 \\
\end{tabular}
\end{ruledtabular}
\smallskip
{\footnotesize  $^{\ddagger}$Experimental ground state.}

\end{table*}

\section{Computational cost estimation}
\label{app:Cost estimation}

\begin{table*}[t]
  \caption{Total number of electronic steps (step 1 $+$ step 2)
    required in the EOS workflow at the sampled volume closest to
    the BM minimum of each functional ($V_\mathrm{BM}$) and at the
    experimental volume ($V$) as provided in the MAGNDATA entry,
    cf.~Table~\ref{tab:exp lattice}.  All counts refer to static
    single-point runs.  For reference, the last two
    columns repeat the cost factors of Table~\ref{tab:cost summary}:
    (A) the cost per SCF step, i.e.\ the median wall time per
    electronic step in the controlled benchmark, with the standard
    deviation over the 10 repetitions in parentheses in units of
    the last digit, and (B) the time to solution, i.e.\ the median
    over all materials and both volume points of the total CPU
    core time ($t\,N_\mathrm{MPI}$), with the median absolute
    deviation in parentheses, each relative to the corresponding
    PW92 run.}
  \label{tab:no scf total}
  \begin{ruledtabular}
  \begin{tabular}{lcccccccccc}
       & \multicolumn{2}{c}{\MnIr}  & \multicolumn{2}{c}{\MnGe} & \multicolumn{2}{c}{\NiS} & \multicolumn{2}{c}{\YMO} & (A) & (B)\\
       & $V_\mathrm{BM}$ & $V$ & $V_\mathrm{BM}$ & $V$ & $V_\mathrm{BM}$ & $V$ & $V_\mathrm{BM}$ & $V$ & cost per SCF step & time to solution\\
    \midrule
    PZ & 151 & 86  & 75 & 34 & 33 & 425 & 28 & 26 & 0.931(5) & 1.4(4)\\
    PW92 & 35 & 33  & 74 & 47 & 34 & 101 & 28 & 26 & 1 & 1\\
    PBE & 31 & 50  & 94 & 110 & 75 & 68 & 104 & 76 & 1.005(5) & 1.5(7)\\
    SCAN & 48 & 51  & 47 & 56 & 123 & 123 & 146 & 232 & 4.62(2) & 9(6)\\
    NCMSCAN & 38 & 45  & 53 & 41 & 65 & 50 & 173 & 172 & 6.96(15) & 7(3)\\
    LFNCBR89-NCCS & 54 & 35  & 31 & 30 & 37 & 37 & 76 & 30 & 6.92(3) & 5(2)\\
  \end{tabular}
  \end{ruledtabular}
\end{table*}

This appendix provides further details on the two perspectives on the computational cost introduced in Sec.~\ref{subsec:Computational cost}.

Procedure for (A): The cost per SCF step is measured in a controlled benchmark, namely \MnIr\ in the experimental cell, with 10 repetitions of 10 \texttt{ALGO}$=$\texttt{Damped} steps restarted from a 5-step Davidson \texttt{WAVECAR}, no wavefunction or charge IO, a single exclusive node with 32 MPI ranks, and a common plane-wave cutoff of 800\,eV for all functionals. The reported cost factor is the median wall time per electronic step relative to the PW92 repetition of the same time block, with the standard deviation over the 10 repetitions given in parentheses.

Procedure for (B): Table~\ref{tab:no scf total} presents the number of SCF steps required to achieve electronic convergence in the EOS calculations described in the last paragraph of Sec.~\ref{subsec:Volume relaxation}. It is an indicator of how smoothly the SCF loop can find a local minimum given a certain XC functional. We see some outliers, where a calculation took significantly longer to converge for a certain combination of system, volume, and functional, e.g., PZ at $V$ for \NiS. The number of steps required for convergence is particularly important for the total computational cost: the overhead of computing the extra densities and terms in the XC potential can be entirely compensated if fewer iterations are necessary to reach a minimum. Column (B) of Table~\ref{tab:no scf total} therefore presents the median of the total core time relative to PW92, which ensures that the outliers do not dominate the interpretation.

\bibliography{references}

\end{document}